\documentclass[jgrga]{AGUTeX}

\authorrunninghead{HEYMAN ET AL.}
\titlerunninghead{SPATIAL CORRELATIONS IN BED LOAD}
 
\authoraddr{Corresponding author: J. Heyman, EPFL - LHE Station 18, 1015 Lausanne. (joris.heyman@epfl.ch)}

\usepackage{epsfig}
\usepackage{amsmath}
\usepackage{amsfonts}

\begin{document}

\title{Spatial correlations in bed load transport: evidence, importance, and modelling.}
\authors{J. Heyman\altaffilmark{1}, H.B. Ma\altaffilmark{2}, F. Mettra\altaffilmark{1} and C. Ancey\altaffilmark{1}}
\altaffiltext{1}{Laboratory of Environmental Hydraulics,  School of Architecture, Civil and Environmental Engineering, \'Ecole Polytechnique F\'ed\'erale de Lausanne, Switzerland}
\altaffiltext{2}{State Key Laboratory of Hydroscience and Engineering, Tsinghua University, Beijing, China}

\begin{abstract}
This article examines the spatial {dynamics of bed load particles} in water. We focus particularly on the fluctuations of particle activity, which is defined as the number of moving particles per unit bed {length}. 
Based on a stochastic model recently proposed by \citet{Ancey2013}, we derive the second moment of particle activity analytically; that is the spatial correlation functions of particle activity. From these expressions, we show that large  moving particle clusters can develop spatially. Also, we provide evidence that fluctuations of particle activity are scale-dependent. Two characteristic lengths emerge from the model: a saturation length $\ell_{sat}$ describing the length needed for a perturbation in particle activity to relax to the homogeneous solution, and a correlation length $\ell_c$ describing the typical size of moving particle clusters. A dimensionless P\'eclet number can also be defined according to the transport model. Three different experimental data sets are used to test the theoretical results. We show that the stochastic model describes spatial patterns of particle activity well at all scales.  In particular, we show that $\ell_c$ and $\ell_{sat}$ may be relatively large compared to typical scales encountered in bed load experiments (grain diameter, water depth, bed form wavelength, flume length...) suggesting that the spatial fluctuations of particle activity have a non-negligible impact on the average transport process. 
\end{abstract}

\begin{article}
\section{Introduction}
Originating in the late 1930s with the seminal work of Hans Albert Einstein \citep{Einstein1937,Einstein1950}, the probabilistic approach to bed load transport has had a surge of interest among the scientific community in recent years \citep{papa2002,Jerolmack2005(2),Ancey2006,ANCEY2008,Valyrakis2010,ANCEY2010,Furbish2013}. This revival has been combined with a substantial improvement in laboratory measurement techniques. In particular, the use of high-speed videos of particle motion together with powerful digital processing, has allowed for ground-breaking precision in the description of sediment particle dynamics \citep{ANCEY2004,Radice2009,Lajeunesse2010,Furbish2012(2),Martin2012}.

{These data allow for an improved understanding of the transport process and its fluctuations. Indeed, bed load transport rates are known to show fluctuations often larger than the mean \citep{Drake1988,kuhnle,hoey1992,Ancey2006,singh2009}. The problem arising in any system exhibiting internal fluctuations is the calculation of consistent average values, or relationships, that can be used to describe its macroscopic behavior.}

{This paper is concerned with drawing possible links between the microscopic stochastic motion of bed load particles and macroscopic variables, such as the average bed load flux or the particle activity (the number of moving particles per unit bed length). In other words, the question we try to answer here is how individual particle motion is reflected through larger scale transport relations. More precisely, how does noise, intrinsically present at small scales, modify the average macroscopic equilibrium at large scales? Until now, the variability of bed load flux has been deliberately ignored in most transport models. Is this approximation physically justified or, on the contrary, do models need to take into account bed load rate fluctuations in order to accurately predict and quantify sediment budgets?}

Among the {recent stochastic models of bed load transport,  \citet{sun2000} proposed a two states Markov model suggesting that bed load transport rates would follow a binomial distribution. \citet{wu2003} considered the rolling and lifting probabilities of particles in a turbulent stream while in \citet{wu2004}, they proposed a stochastic partial transport model for mixed size sediments.} \citet{turowski2010} suggested that the shape of the probability distribution of the bed load flux was a function of the inter-arrival time of particles. More recently, in four companion papers, \citet{Furbish2012(1)} provided further insights into particle random motion and its consequences on macroscopic conservation equations. Using an ensemble averaging procedure, they found that the bed load flux comprises both an advective and a diffusive term due to particle velocity fluctuations. \citet{ANCEY2008} developed a stochastic erosion/deposition model describing the fluctuation of the number of moving particles in an observation window. Using the framework of birth-death Markov processes, they provided a comprehensive picture of the large fluctuations observed in their experiments.  Generalizing \citeauthor{ANCEY2008}{{'s}} \citeyearpar{ANCEY2008} probabilistic model, \citet{Ancey2013} were able to model the spatial variability of particle activity. By studying the erosion, deposition and motion of particles on a lattice made of regular cells, they ended up with a stochastic equation describing the process in both space and time. The model is valid for low to moderate transport rates.

In this paper, we explore some applications of the stochastic model recently proposed by \citet{Ancey2013}. While \citet{Ancey2013} paper concerned the theoretical foundations of the model, this article focuses on validation issues. In doing so, we will demonstrate how the model accurately reproduces spatial fluctuations of the bed load particle activity. {This will be achieved by comparing theoretical results with various experimental data of particle trajectories {in time and space} (Fig.~\ref{lattice}).} 

The paper is organized as follows. First, to make the article self-contained, we briefly go over how the general stochastic equations governing the bed load phase are derived. A rigorous and detailed derivation is not provided here since it is available in \citet{Ancey2013}. Then, we compute the second moment (the spatial correlation function) of the particle activity. {We also give the analytical expression of the $K$-function \citep{RipleyK}, often used in point process analysis to highlight the possible correlations existing between particle locations.} In the last section, we use three different experimental studies to test the model, two of which have already been published \citep{ANCEY2004,Furbish2012(2)}. {The third study is an original data set which consists of three experiments carried out in a steep slope flume.} A general method to calibrate model parameters on experimental data is proposed. {Finally, we discuss the importance of including spatial variability in bed load transport models and we propose possible improvements of \citeauthor{Ancey2013}{{'s}} \citeyearpar{Ancey2013} stochastic model.}

\section{Theory}
\subsection{Physical space}
The transport of bed load particles occurs in a thin layer over the surface of an erodible bed. Particles generally move in a preferential direction (down the slope, parallel to fluid flow) so that it is possible to {constrain} the study to a one-dimensional space in that principal direction. A generalization to a two-dimensional space, while technically possible, goes beyond the scope of this paper.

\begin{figure}
\includegraphics[width=\linewidth]{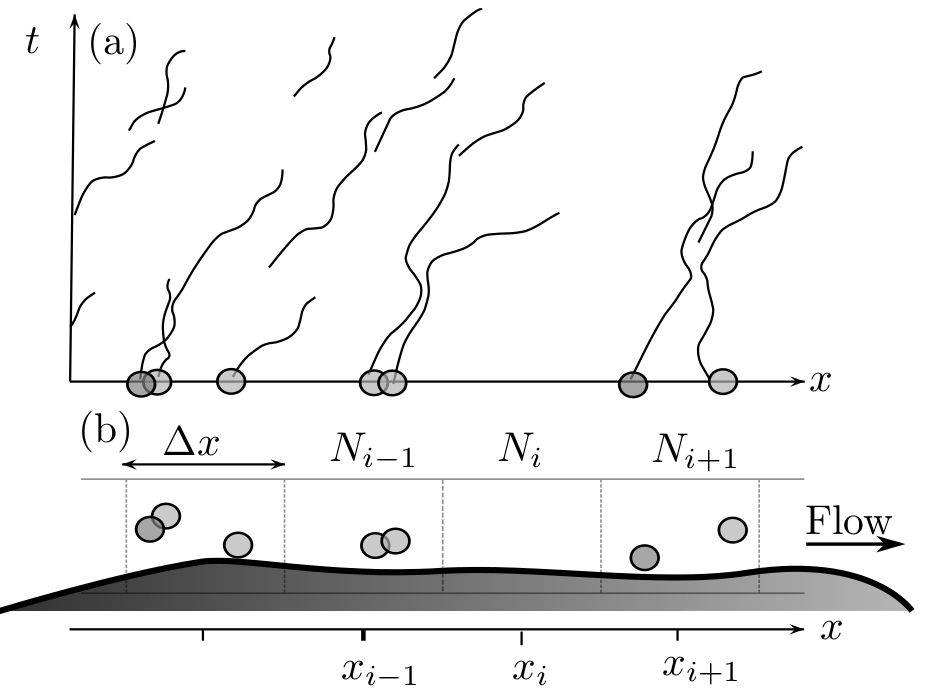}
\caption{(a) Particle trajectories in a time-space plane. (b) Discretization of the space in cells of equal length ($\Delta x$) and number of particles in each cell $i$ at a given time $t$ ($N_i$).}
\label{lattice}
\end{figure}

Let us consider a one-dimensional space that represents a river reach, or an experimental flume. 
The space is divided into cells of equal length $\Delta x$ {(Fig.~\ref{lattice})}. Each cell of this lattice is labelled by an index $i$. We call $N_i(t)$ the random variable describing the number of moving particles $n_i$ in cell $i$ at time $t$.  We introduce the multivariate probability
\begin{equation}
P\left(\left[n_1,n_2,\dots \right],t\right)=P(\boldsymbol{n},t), 
\end{equation} 

where $\boldsymbol{n}$ is the vector of all $n_i$. In other words, $P(\boldsymbol{n},t)$ is the probability of simultaneously observing $N_1(t)=n_1, N_2(t)=n_2, \cdots$ at time $t$. The particle activity in cell $i$ is defined as
\begin{equation}
\gamma(x_i,t)=N_i(t)/\Delta x,
\end{equation} 

where $x_i$ denotes the position of the center of the cell $i$.

\subsection{Particle motion}
Bedload transport describes the motion of bed particles (sliding, rolling or saltating) sheared by a fluid. {To build their model,} \citet{ANCEY2008} and \citet{Ancey2013} distinguished three independent {phases} of particle motion: entrainment, transport and deposition. {Those are briefly reviewed below.}

The entrainment of a resting particle by a fluid flow has been extensively studied and its intermittent and random character is widely accepted \citep{Einstein1950,papa2002,wu2003,Schmeeckle2007,DETERT,celik2010,Valyrakis2010,dwivedi2011}. {For flow conditions close to incipient sediment motion, the fluid flow intermittently dislodges particles from the bed. Turbulent flow structures being spatially and temporally correlated, it is also likely that several particles are entrained simultaneously, leading to clouds of moving particles \citep{nelson2005,Drake1988}. Various experiments suggested that different mechanisms of entrainment exist, such as entrainment caused by a particle collision or by a local bed rearrangement \citep{Schmeeckle2001,Heyman2013}.}

{In their model, \citet{ANCEY2008} conceptualized the entrainment of a particle as a sum of two basic random processes: (i) a memoryless and uncorrelated process, referred to as entrainment, and (ii) a correlated process with intensity proportional to the number of particles already in motion, and referred to as collective entrainment. The probability of a particle being entrained ($n_i \rightarrow n_i+1$) in a cell $i$ of length $\Delta x$ during a small time interval $\mbox{d} t$ is thus}
\begin{equation}
P(n_i \rightarrow n_i+1)=(\lambda \Delta x + \mu n_i)\mbox{d}  t,
\end{equation} 

{where $\lambda$~[particles~m$^{-1}$~s$^{-1}$] is the mean entrainment rate of particles per unit length and $\mu$~[s$^{-1}$] is the collective entrainment rate.}

{After being entrained, a particle is dragged by the fluid flow for a certain time before depositing onto the bed. \citet{ANCEY2008} envisioned the deposition of a particle as a memoryless and independent random process. Thus, at any time, the probability of observing a single particle deposition ($n_i \rightarrow n_i-1$) in the cell $i$ during $\mbox{d}  t$ can be expressed as}
\begin{equation}
P(n_i \rightarrow n_i-1)=\sigma n_i \mbox{d}  t,
\end{equation} 

{where $\sigma$~[s$^{-1}$] is the mean particle deposition rate. Note that, in this basis, $1/\sigma$ is the mean travel time of a particle.}

{Once put in motion, particles are transported downstream by the fluid flow. Their velocity is frequently altered due to repeated impacts on the bed as well as by drag fluctuations due to turbulence. \citet{Ancey2013} proposed a model of Brownian motion in a potential to describe the transport process. They found that, under certain conditions, the transport of particles could be described locally by the sum of two contributions: (i) a deterministic advection at the average particle velocity $\bar{u}_s$ and (ii) a random jump process between lattice cells that can be described by the following transition probabilities in the infinitesimal time interval $\mbox{d}  t$:}

\begin{eqnarray}
P(n_i \rightarrow n_i-1,n_{i-1}\rightarrow n_{i-1}+1 )&=&d n_i \mbox{d}  t,  \nonumber \\
P(n_i \rightarrow n_i-1,n_{i+1}\rightarrow n_{i+1}+1 )&=&d n_i \mbox{d}  t,
\end{eqnarray} 

{where $d$ [s$^{-1}$] is a local diffusivity rate. In the following statistical analysis, we do not consider the contribution of advection in the transport of particles since it is deterministic and is not contributing to the fluctuations of the number of moving particles \citep{Ancey2013}. Naturally, the advective contribution to the particle transport process will be reintroduced later in the deterministic part of the equations.}

\subsection{Birth-death process and Poisson representation}
In the previous section, we implicitly assumed that the rate coefficients $\lambda,\sigma,\mu$ and $d$ were constant in space and time.
{This leads us to focus on fluctuations that precisely originate from the randomness of particle motions and exchanges with the bed rather than fluctuations arising because of local changes in flow or bed slope  ---that would in turn modify the rate coefficients, when bedforms are present for instance.}

The transition probabilities defined above form the elementary rules governing the evolution of a multivariate birth-death Markov process (i.e.: a memoryless process of many variables, which evolves by unitary jumps).  From these simple rules, \citet{Ancey2013} derived the multivariate master equation describing the temporal evolution of $P(\boldsymbol{n},t)$,

\begin{eqnarray}
\label{eq:master-total}
\frac{ \partial P(\boldsymbol{n}, t)}{\partial t}  &=& {{\sum}}_i  d (n_i+1)\left(P( \boldsymbol{n}+\boldsymbol{r}_{i}^++\boldsymbol{r}_{i+1}^-,t)+ P(\boldsymbol{n}+\boldsymbol{r}_{i}^++\boldsymbol{r}_{i-1}^-,t)\right) \\ \nonumber
&+&(n_i+1)\sigma P( \boldsymbol{n}+\boldsymbol{r}_i^+   ,t)+\left(\lambda \Delta x+(n_i-1)\mu \right)P( \boldsymbol{n}+\boldsymbol{r}_i^-,t) \\
&-& \left(\lambda \Delta x+n_i(\sigma+\mu)+2 d n_i\right)P(\boldsymbol{n}, t), \nonumber
\end{eqnarray} 

{where $\boldsymbol{r}_i^{\pm}$ is a vector whose elements are all zeros except for its $i$-th value: $r_i=\pm 1$, $r_k=0$ for $k\neq i$. $P( \boldsymbol{n}+\boldsymbol{r}_{i}^++\boldsymbol{r}_{i-1}^-,t)$ is thus the probability of observing the system in the state $\boldsymbol{n}'=(n_1, n_2, \ldots, n_{i-1}-1, n_i+1,n_{i+1}, \ldots)$.}

{This master equation can be greatly simplified using the Poisson representation. Similarly to Laplace or Fourier transforms in the spectral theory of time series, the Poisson representation is a linear operator that transforms a discrete probability space into a continuous one \citep{Gardiner1977}}. {More precisely, it assumes that the probability function of $N_i$ can be decomposed into Poisson distributions with various rates $a_i$,}
 \begin{equation}\label{poissrep}
P(n_i,t)= \int_{\mathcal R^+}\frac{e^{- a_i}a^n_i}{n!}f(a_i,t)\mbox{d}  a_i.
 \end{equation}

{Since the only parameter of a Poisson distribution is also its mean, $a_i$ can be interpreted as the mean number of particles in the cell $i$. On the other hand, $f(a_i,t)$ is the probability of observing the Poisson rate $a_i$ in cell $i$ at time $t$.}

{By inserting Eq.~$\eqref{poissrep}$ in the master equation $\eqref{eq:master-total}$, \citet{Ancey2013} showed that $f(\boldsymbol{a},t)$ ---$\boldsymbol{a}$ being the vector of all $a_i$--- follows an explicit Fokker--Planck equation (i.e., a partial differential equation governing the time evolution of probability functions)}

\begin{eqnarray}
\label{eq:multi-f}
\frac{\partial  }{\partial t}f(\boldsymbol{a},t)&=&\sum_i\mu\frac{\partial^2 a_i f(\boldsymbol{a},t)  }{\partial a^2_i} \nonumber\\
&+& \quad \frac{\partial  }{\partial a_i}\left[f(\boldsymbol{a},t) \left(\lambda \Delta x-a_i(\sigma-\mu)\right)\right] \nonumber\\
&+& \quad \frac{\partial  }{\partial a_i}\left[f(\boldsymbol{a},t) d(a_{i+1}+a_{i-1}-2a_{i }) \right].
\end{eqnarray}

Equivalently, $a_i$ can be shown to follow a {Langevin stochastic equation (i.e., a differential equation with both a deterministic and a stochastic parts)}

\begin{eqnarray}
\mbox{d}  a_i(t)&=&\left( d(a_{i+1}+a_{i-1}-2a_{i })+ \lambda\Delta x-a_i(\sigma-\mu) \right)\mbox{d}  t \nonumber \\
&+&\sqrt{2\mu a_i}~ \mbox{d}  W_i(t),
\label{anceyB}
\end{eqnarray} 

{where $\mbox{d}  W_i(t)$ is the derivative of a Wiener random process, which may be interpreted as a time uncorrelated noise (also called white noise). This noise is said to be multiplicative since its intensity is modulated by $\sqrt{a_i}$ (in contrast to an additive noise which is independent of the state of the process, as it was for example assumed in \citet{Jerolmack2005(2)}).}

Just like the definition of the particle activity $\gamma(x,t)$, let us call $\eta(x,t)$ the Poisson rate per unit bed length (referred to as Poisson activity in the following). We have
\begin{equation}
\eta(x_i,t)=a_i(t)/\Delta x.
\label{poissactivity}
\end{equation}

Using Eq.~$\eqref{anceyB}$ and letting $\Delta x\rightarrow 0$, we obtain the Langevin stochastic partial differential equation for the Poisson activity

\begin{eqnarray}
\mbox{d}\eta(x,t)&=&\left[D\nabla^2 \eta(x,t) + (\mu-\sigma)\eta(x,t)+\lambda\right]\mbox{d}t \\
&+&\sqrt{2\mu\eta(x,t)}\mbox{d}W(x,t), \nonumber
\label{poiss1d}
\end{eqnarray} 

where $W(x,t)$ is now a spatial Wiener process satisfying the condition
\begin{equation}
\mbox{d}W(x,t)\mbox{d}W(x',t) =\delta(x-x')\mbox{d}  t. 
\label{eq:poiss1d2}
\end{equation} 

According to Eq.~$\eqref{eq:poiss1d2}$, the multiplicative noise term $\sqrt{2\mu\eta}\mbox{d}W$ arising in Eq.~$\eqref{poiss1d}$ is perfectly uncorrelated in space and time. We also introduced the notation $D=d \Delta x^2$, {which highlights the connection between} the local particle jump rate $d$ [s$^{-1}$] and the macroscopic particle diffusivity $D$ [m$^2$~s$^{-1}$]. 

{Making use of the linearity of the deterministic part of  Eq.~$\eqref{poiss1d}$, we can reintroduce the deterministic advection flux}

\begin{eqnarray}
\mbox{d}\eta(x,t)&=&\left[- \bar{u}_s \nabla \eta(x,t) + D \nabla^2 \eta(x,t) \right] \mbox{d}t  \nonumber \\
&+&\left[\lambda-(\sigma-\mu)\eta(x,t)\right]\mbox{d}t    \\
&+&\sqrt{2\mu\eta(x,t)} \mbox{d}W(x,t) \nonumber
\label{lang2}
\end{eqnarray} 

Eq.~$\eqref{lang2}$ models the stochastic evolution of the rate (per unit length) of the Poisson distribution followed by $N_i(t)$. It is shown in Appendix~\ref{app:PP} how Eq.~$\eqref{lang2}$ can be solved numerically and how it can be related to the point process framework~\citep{coxbook}.  Eq.~$\eqref{lang2}$ also shares interesting similarities with the BCRE model of dry granular avalanches of \citet{bouchaud1995}. In Appendix~\ref{app:BCRE}, we show how Eq.~$\eqref{lang2}$ may be used as a stochastic version of the BCRE model in order to characterize spatial correlations in some dry granular flows.

\subsection{Moments}
In the remaining of the paper, the notation $\left\langle \bullet \right\rangle$ denotes ensemble averaging (i.e., average over all the possible states of a stochastic process). There exists a simple connection between moments of $a$ in the Poisson representation and moments of the real variable $N$.  Indeed, the $p$-factorial moment of $N$ (i.e., $\left\langle n(n-1)(n-2)\dots\right\rangle$) is equal to the $p$-moment of $a$ (i.e., $\left\langle a^p \right\rangle$), implying that $\left\langle  n  \right\rangle =\left\langle  a \right\rangle $ and $\left\langle  n^2  \right\rangle = \left\langle  a^2 \right\rangle +\left\langle  a  \right\rangle$ \citep{Gardiner1977,Ancey2013}. Similar relationships exist between between moments of $\eta$ and moments of $\gamma$: $ \left\langle \eta(x,t) \right\rangle = \left\langle \gamma(x,t) \right\rangle$ and $\left\langle \gamma(x,t),\gamma(x',t)\right\rangle=\left\langle{\eta(x,t),\eta(x',t)}\right\rangle+\delta(x-x')\left\langle{\eta(x,t)}\right\rangle$ \citep{Gardiner1977}.   In the following, we study the first and second moments of Eq.~$\eqref{lang2}$.

\subsubsection{First moment}
The average behavior of $\eta(x,t)$ is easily obtained by dropping the noise term in Eq.~$\eqref{lang2}$,
\begin{equation}
\frac{ \partial \left\langle \eta \right\rangle}{\partial t}+ \bar{u}_s \frac{ \partial \left\langle \eta \right\rangle}{\partial x}= D \frac{ \partial^2 \left\langle \eta \right\rangle}{\partial x^2} +\lambda-(\sigma-\mu)\left\langle \eta \right\rangle.
\label{lang2av}
\end{equation} 

It is a linear advection-diffusion-reaction equation. For $t\to\infty$, providing that $\sigma>\mu$, the stationary and homogeneous solution is
\begin{equation}
\left\langle \eta \right\rangle_s =\frac{\lambda}{\sigma-\mu},
\label{eq:average}
\end{equation} 

where the notation $\left\langle  \bullet \right\rangle_s$ denotes the ensemble average for stationary and homogeneous conditions. Thus, the stationary homogeneous particle activity is $\left\langle \gamma \right\rangle_s = \lambda/(\sigma-\mu)$.

\begin{figure}
\centering
\includegraphics[width=\linewidth]{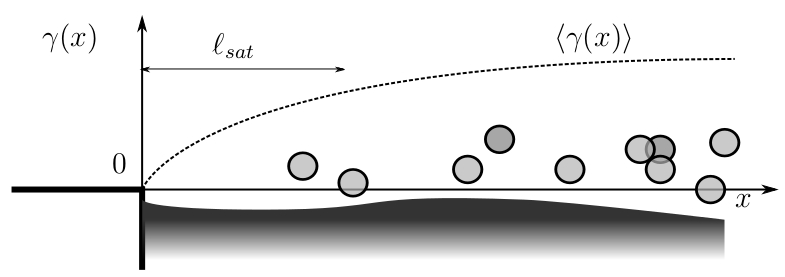}
\caption{Boundary value problem corresponding to the relaxation of particle activity to equilibrium. $\ell_{sat}$ is the saturation length.}
\label{im:sat}
\end{figure}

{\citet{Ancey2013} studied the evolution of a sediment pulse in space and time. Here, we focus on the stationary behavior of $\eqref{lang2av}$ given a fixed Dirichlet boundary condition at the origin $\left\langle \gamma(0) \right\rangle=0$ (Fig.~\ref{im:sat}).
}
{The problem simplifies to a second order ordinary differential equation whose solution is given by}

\begin{eqnarray}
\left\langle \gamma(x) \right\rangle &=& \frac{\lambda}{\sigma-\mu}\left(1-e^{-x/\ell_{sat}}\right), \\
 \ell_{sat} &=&\frac{2 \ell_c}{{\mbox{Pe}}}\left(\sqrt{1+4{\mbox{Pe}}^{-2}}-1\right)^{-1}, \nonumber
 \label{eq:satPe}
\end{eqnarray} 

{where we have introduced $\ell_c=\sqrt{D/(\sigma-\mu)}$ and the dimensionless number ${\mbox{Pe}}={\bar{u}_s \ell_c}/{D}$, which can be interpreted as a local P\'eclet number. The P\'eclet number is usually defined as the ratio of the rate of advection by the flow to the rate of diffusion. In the case of the diffusion of matter, it can also be defined as the product of a typical length scale by the advection velocity divided by the diffusivity. In our case, the local P\'eclet number compares particle diffusion against advection with respect to the correlation length $\ell_c$. The latter originates from the coupled action of diffusion and particle exchanges with the bed (collective entrainment and deposition) but its meaning will be best understood while studying spatial correlations. Note that, in contrast to aeolian sediment transport, the saturation length $\ell_{sat}$ does not originate from particle inertia (which is negligibly small in water) but rather from the particle exchanges with the bed and their transport by the flow \citep{charru2006}. }

{Unfortunately, no experimental data about saturation length in bed load transport under water are presently available, so that no comparison could be made with the theoretical predictions.}

\subsubsection{Second moment}

We show in Appendix~\ref{app:space} that the stationary and homogeneous spatial correlation function of the particle activity reads

\begin{eqnarray}
&& \left\langle \gamma(x),\gamma(x')\right\rangle_s \equiv \left\langle \gamma(x)\gamma(x')\right\rangle_s - \left\langle \gamma(x)\right\rangle_s^2\\
 &&= \delta(x-x')\left\langle{\gamma}\right\rangle_s+\frac{ \left\langle{\gamma}\right\rangle_s \mu}{2 \ell_c(\sigma-\mu)}\exp{\left(-\frac{|x-x'|}{ \ell_c} \right)} \nonumber ,
\label{coreff}
\end{eqnarray} 

where we have already used the correlation length $\ell_c=\sqrt{D/(\sigma-\mu)}$. The steady-state spatial correlation function is thus the sum of a Dirac delta function (i.e., $\delta(x)$ is a distribution which is zero everywhere except at $x=0$, with an integral of one over the entire real line) of intensity equal to the mean density of moving particles and an exponentially decaying function corresponding to correlations caused by the collective entrainment of particles (when $\mu=0$, this term disappears). The correlation length $\ell_c$ modulates the speed of the decay. Thus, $\ell_c$ is a measure of the typical spatial scale of fluctuations in particle activity. It increases with the diffusivity of particles and with the collective entrainment rate, but decreases with the deposition rate. Interestingly enough, when $\mu=0$, no more spatial correlations are observed in the particle activity although $\ell_c$ remains positive. When $\mu \rightarrow \sigma$, the spatial correlations of particle activity become infinitely large and when $\mu=\sigma$, Eq.~$\eqref{lang2}$ becomes unstable and an exponential increase in the number of moving particles is observed. 

Another quantity of interest, often used to describe a spatial point process, is the conditional intensity $h(x-x')$, which gives the conditional probability of finding a particle at $x'$ given that there is a particle at $x$ \citep{coxbook}. The conditional intensity and the correlation function are directly related by
\begin{equation}
\left\langle \gamma(x,t),\gamma(x',t)\right\rangle_s=\delta(x-x')\left\langle{\gamma}\right\rangle_s + \left\langle{\gamma}\right\rangle_s h(x-x') - \left\langle{\gamma}\right\rangle_s^2, \nonumber
\end{equation} 

so that by identification, we have
\begin{equation}
h(x-x')= \left\langle{\gamma}\right\rangle_s+\frac{1}{2 \ell_c}\frac{\mu}{\sigma-\mu}\exp{\left(-\frac{|x-x'|}{ \ell_c} \right)}.
\end{equation} 

A more convenient function for data analysis is the $K$-function \citep{RipleyK}, where $K(x)$ represents the expected number of moving particles found in a ball of radius $x$ centered on a particle location, divided by the mean process rate. This can be calculated from the conditional intensity function by

\begin{eqnarray}
K(x)&= &\frac{1}{\left\langle{\gamma}\right\rangle_s}\int_{0}^{x} h(u) \mbox{d}  u \nonumber \\
 &=& x +  \frac{1}{\left\langle{\gamma}\right\rangle_s}\frac{\mu}{\sigma-\mu}\left[1-\exp{\left(-\frac{x}{ \ell_c} \right)}\right]. 
 \label{Kfunc}
\end{eqnarray} 

In the case of a Poisson point process in one dimension (i.e., a process which is spatially uncorrelated),  $K(x)=x$. Furthemore, Eq.~$\eqref{Kfunc}$ shows that $K(x)>x$ if $\mu>0$; so the point process formed by particle locations is said to be clustered (see Appendix~\ref{app:PP}). 
\section{Applications}

After deriving the spatial correlation function of the particle activity, we examine how it can be compared with experimental data.

\subsection{Experiments}
\begin{figure*}
 \centerline{\includegraphics[width=17cm]{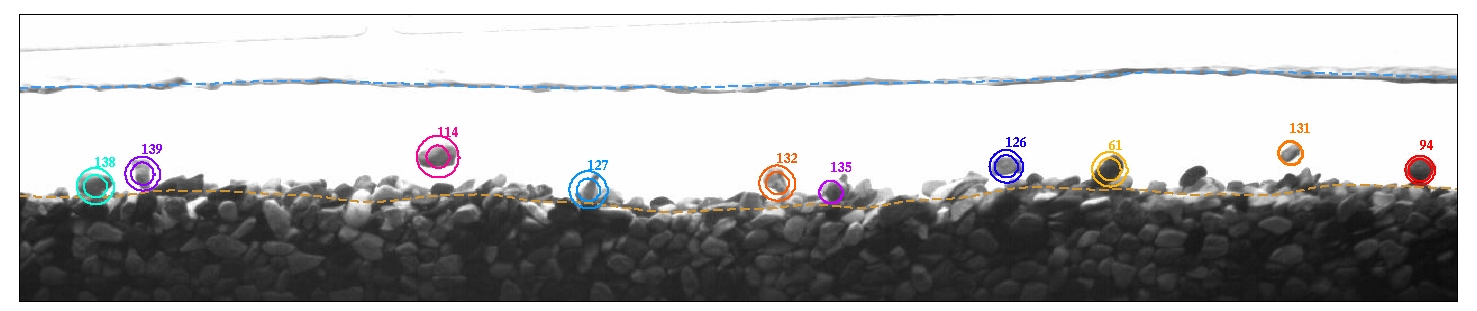}}
 \caption{Snapshot from one of the two cameras used in experiments J (the field of vision is about 50~cm long). Moving particle locations are estimated by the tracking algorithm (circles show the confidence interval of the particle diameter estimates, while numbers refer to the trajectory index $j$). The water surface and the bed elevation are also detected automatically (blue and orange dashed lines).}
 \label{newexp}
 \end{figure*} 
We use three different experimental data sets. Two of them have been previously published \citep{ANCEY2004,Furbish2012(2)}. The third comes from an experimental setup especially built by the authors to observe spatial and temporal fluctuations of bed load transport (Fig.~\ref{newexp}). All three studies provide high resolution measurements of particle trajectories using high speed videos. Details of the experimental setups are given in Appendix~\ref{app:setups} while experimental conditions are reported in Table~\ref{tab:table1}.

Hereafter, we denote all \citet{ANCEY2004} experiments by using the prefix B, \citet{Furbish2012(2)} experiment using R, and the new data set using J. The numbers following the prefix specify experimental slope and solid discharge. For instance B10-5 stands for \citet{ANCEY2004} experiment conducted using a 10\% sloping flume with a mean solid discharge of 5 particles~s$^{-1}$.  

Typically, an experimental outcome consists of an ensemble of particle locations in the streamwise direction: $x_{j,k}$ is the position of particle $j$ at frame $k$. In total, experiment B gathers more than 8000 particle trajectories over 4 minutes, experiment R gathers more than 300 particles trajectories over about 0.4~s while each experiment J gathers in average 5000 trajectories over 10 minutes.

\begin{table*}
\begin{center}
\def~{\hphantom{0}}
 \scriptsize  
\begin{tabular}{l|ccccccccc|ccccc|ccccccccccc}
  &$B$& $d_{50}$ & $\tau_s$ & Fr & tan$\theta$ & $\bar{v}$ & $\bar{h}$&$\bar{q}_s$& $\bar{\gamma}$&$\bar{u}_s$&$\sigma$&$D$&$\lambda$&$\mu$ &$\ell_c$&${\mbox{Pe}}$&$I(\infty)$&$\ell_{sat}$\\ 
\hline
\hline
B10-5 & 0.6 & 6 & 0.11 & 1.42  & 10.0 & 0.41 & 1.0 & 4.9 & 26.9 &  0.170 & 2.72 & 15 &  24 & 1.825 & 4.1 & 4.6 & 3.0 & 20 \\
\hline
R0-79 & 6.0 & 0.5 & 0.06 & 0.35 & - & 0.31 & 12.5 & 78.7 & 1711.0 & 0.046 & 1.85 & 1.5  & 171 &1.754 & 3.8& 12.2& 18.6 & 47 \\ 
\hline
J3-1 & 3.5 & 7 & 0.14 & 1.30  & 3.5 & 0.80 & 3.8 & 1.4 & 4.6 &  0.310 & 0.52 & 59  & 0.33 & 0.447 & 28.7 & 15.1 & 7.2 & 434  \\ 
J4-1 & 3.5 & 7 & 0.17 & 1.39 & 4.5 & 0.86 & 3.9 & 1.2 & 3.9 & 0.300 & 0.50 & 89  & 0.39 & 0.403 & 29.8 &10.1 & 5.0 & 303  \\ 
J5-1 & 3.5 & 7 & 0.14 & 1.47  & 4.7 & 0.80 & 3.1 & 0.9 & 2.8 & 0.320 & 0.56 & 55  & 0.27 & 0.470 & 24.3 & 14.1 & 6.0 & 345 \\
\hline
\end{tabular}
\end{center}
\caption{Experimental parameters and model fits. $B$~[cm], channel width; $d_{50}$~[mm], mean particle diameter; $\tau_s$~[-], Shields stress; Fr~[-], Froude number; $\tan(\theta)$~[\%], slope angle; $\bar{v}$ [m~s$^{-1}$], mean flow velocity; $\bar{h}$~[cm], mean water depth; $\bar{q_s}$~[particles~s$^{-1}$], mean output solid discharge;  $\bar{\gamma}$~[particles~m$^{-1}$], mean activity; $\bar{u}_s$~[m~s$^{-1}$], average particle velocity; $\sigma$~[s$^{-1}$], deposition rate; $D$ [cm$^2$~s$^{-1}$], particle diffusivity; $\lambda$~[particle~m$^{-1}$~s$^{-1}$], particle entrainment rate per meter length; $\mu$~[s$^{-1}$], collective entrainment; $\ell_c$~[cm], correlation length ($\ell_c=\sqrt{D/(\sigma-\mu)}$); ${\mbox{Pe}}$~[-] local P\'eclet number (${\mbox{Pe}}=\ell_c \bar{u}_s/D$); $I(\infty)$~[-], limiting value of the dispersion index; $\ell_{sat}$~[cm] saturation length (Eq.~$\eqref{eq:satPe}$).}
\label{tab:table1}
\end{table*}

\normalsize
\subsection{Spatial fluctuations}
Let us consider the number of moving particles in a window of length $L$ at a given time $t$,
\begin{equation}
N(L,t)= \int_L \gamma(x,t) \mbox{d}x.
\end{equation} 

{When $t\to\infty$, the stationary average of $N(L,t)$ is}

\begin{eqnarray}
\mbox{Mean}[N(L)] &\equiv & \left\langle  \int_L  \gamma(x,t) \mbox{d}x  \right\rangle_s = \int_L  \left\langle \gamma(x,t) \right\rangle_s \mbox{d}x \nonumber \\
&=& \left\langle{\gamma}\right\rangle_s L, 
\end{eqnarray} 

while the variance of $N(L,t)$ (sometimes called the variance of the sample mean) is defined by

\begin{eqnarray}
\mbox{Var}[N(L)]&\equiv & \left\langle   \int_L  \gamma(x,t) \mbox{d}x   \int_L  \gamma(x',t) \mbox{d}x' \right\rangle_s -\left\langle   \int_L  \gamma(x,t) \mbox{d}x \right\rangle_s^2 \nonumber \\
&=& \int_L \int_L \left\langle\gamma(x,t),\gamma(x',t)\right\rangle_s \mbox{d}x~ \mbox{d}x'.
\label{vareq}
\end{eqnarray} 

Introducing Eq.~$\eqref{coreff}$ into Eq.~$\eqref{vareq}$ and integrating it (see Appendix \ref{A4}), we find

\begin{eqnarray}
\mbox{Var}[N(L)]&=& \left\langle{\gamma}\right\rangle_s L\\
&+&\left\langle{\gamma}\right\rangle_s \ell_c \frac{\mu}{\sigma-\mu}\left(L/\ell_c+\mbox{e}^{-L/\ell_c}-1\right). \nonumber
\label{vareq3}
\end{eqnarray} 

Eq.~$\eqref{vareq3}$ shows the dependence of the variance of $N(L)$ on the length $L$ of the observation window. Let us define the dispersion index $I(L)$ as the ratio of the variance over the mean
\begin{equation}
I(\tilde{L})=\frac{\mbox{Var}[N(\tilde{L})]}{\mbox{Mean}[N(\tilde{L})]}= 1+ \frac{\mu}{\sigma-\mu}\left(1+\frac{\mbox{e}^{-\tilde{L}}-1}{\tilde{L}}\right),
\label{varmean}
\end{equation} 

with $ \tilde{L}=L/\ell_c$.

\begin{figure}
\centering
\includegraphics[width=\linewidth]{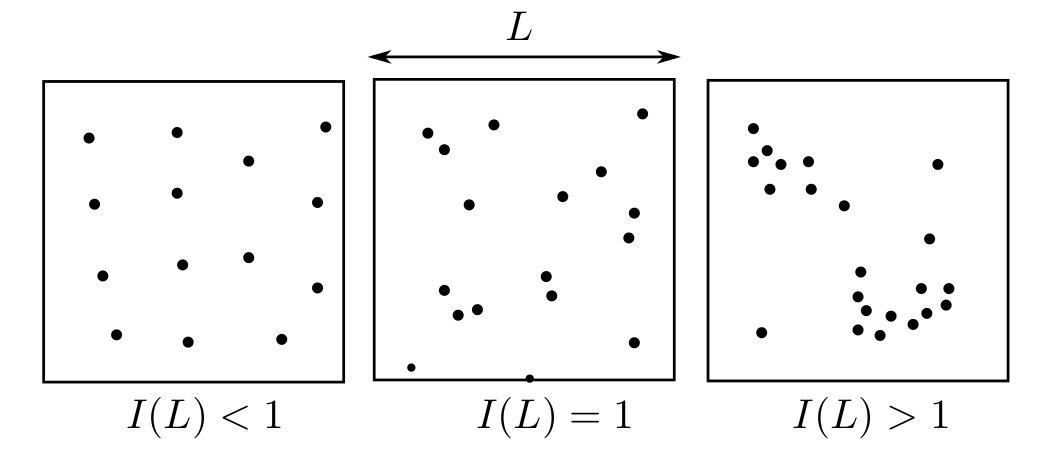}
\caption{Example of a realization of point positions in a two-dimensional space depending on the value of the dispersion index $I(L)$. Here, $L$ can be interpreted as the size of the box.}
\label{indexofdisp}
\end{figure}

The dispersion index is used to characterize the relative positions of points (particle locations). Three classes of stochastic processes are generally distinguished depending on the value of $I$: under-dispersed processes for $I<1$; purely random processes (or Poisson processes) when $I=1$; and over-dispersed or clustered processes when $I>1$ (Fig.~\ref{indexofdisp}).

\begin{figure*}
\centerline{\includegraphics[width=17cm]{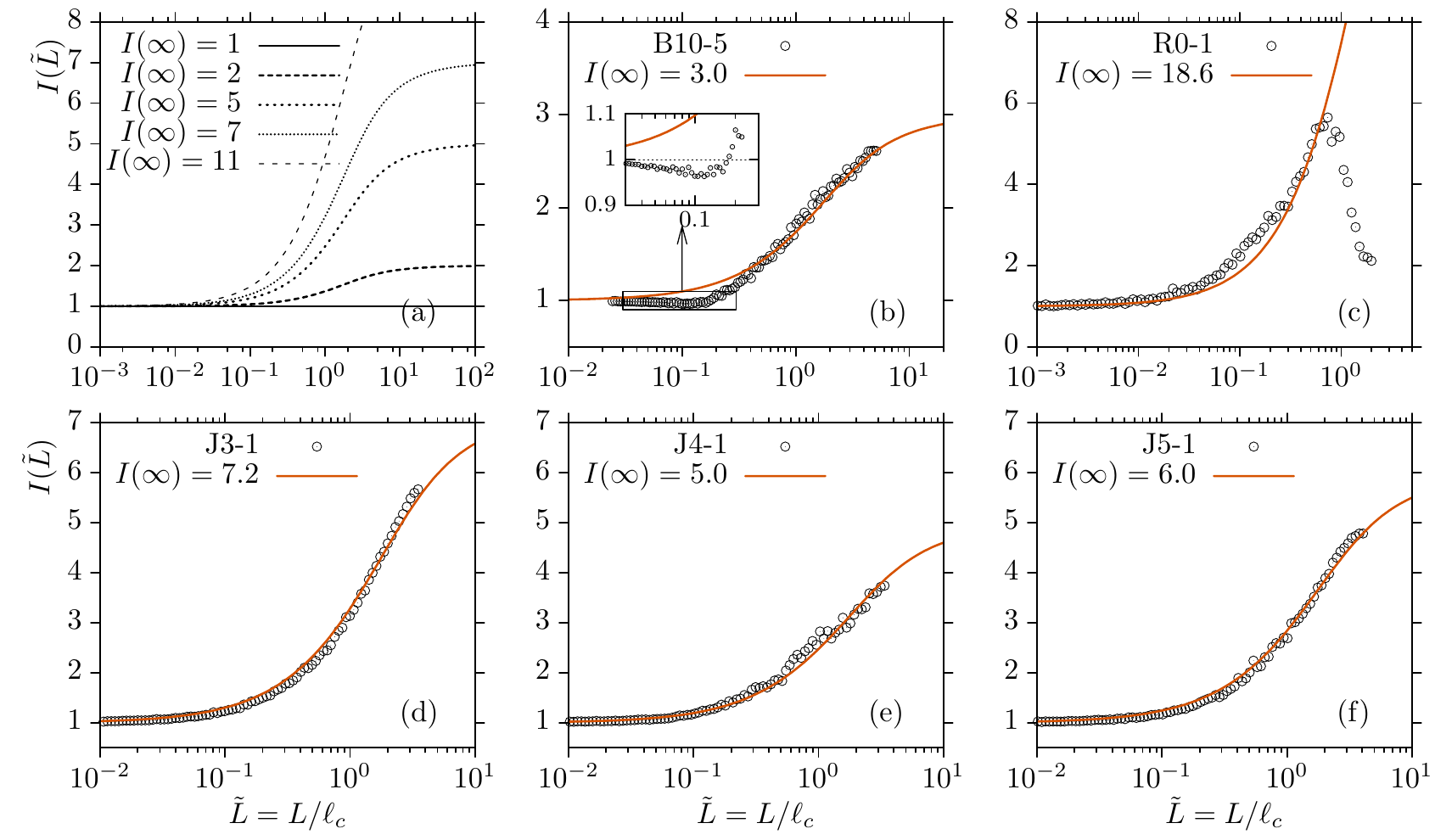}}
\caption{Dispersion index ($I(\tilde{L})$, with $\tilde{L}=L/\ell_c$). (a) Eq.~$\eqref{varmean}$, (b) experiment B, (c) experiment R, (d-e-f) experiments J.}
\label{fig:disp}
\end{figure*}

The theoretical dispersion index $\eqref{varmean}$ grows from one, when the observation window is small, to the constant value $I(\infty)=1+\mu/(\sigma-\mu)$, as the window length tends to infinity (Fig.~\ref{fig:disp}(a)). In other words, depending on the observation scale $L$, the number of moving particles in the observation window exhibits a different statistical behavior.

When $L$ tends to $0$, $I(L)$ tends to one, so that the variance and the mean of $N(L)$ are equal. Thus, in the small scale limit, $N(L)$ tends to a Poisson process. This is mathematically explained by the presence of the Dirac delta function in the spatial correlation function $\eqref{coreff}$. In other words, for decreasing values of $L$, a limit will be reached when most of the time an observation window contains no particle, and more rarely one. {The small $L$ limit can be seen as representative of a Bernoulli process (i.e.,  $N(L)=1$ with probability $\left\langle{\gamma}\right\rangle_s L$ and $N(L)=0$ with probability $1-\left\langle{\gamma}\right\rangle_s L$), which is well approximated by a Poisson process when $\left\langle{\gamma}\right\rangle_s L\to 0$.}

On the contrary, when $L \rightarrow \infty$, $I$ reaches a constant value $I(\infty)$. Note that $I(\infty)>1$ if $\mu>0$, so that the variance of $N(L)$ is now greater than its mean. Thus, for larger scales and when $\mu>0$, $N(L)$ cannot be described anymore by a Poisson process. Thus, moving particles are expected to form clusters during their motion when collective entrainment is considered (Fig.~\ref{indexofdisp}). 

Experimental dispersion indices are presented in Fig.~\ref{fig:disp} (b--f). The procedure used to compute such index is presented in Appendix~\ref{app:disp}. In all experiments, dispersion indices change through spatial scales. From a Poisson type process at small scales ($\tilde{L} \rightarrow 0$), $I(\tilde{L})$ increases with increasing scales (Fig.~\ref{fig:disp}). The dispersion index of experiment R follows a slightly different evolution, since it drops at scales larger than 5~cm. This behavior might be explained by the relatively short measurement window ($\sim 8$~cm) and the relatively small number of frames available in experiment R, which leads to biased estimates of the dispersion index (see Appendix~\ref{app:disp}). 

One striking feature of the experimental dispersion index that appears only in experiment B is the slight decrease below unity for lengths of the order of the particle diameter (Fig.~\ref{fig:disp}(b)). This phenomenon results from negative values in the correlation function at those scales and cannot be described by the Markov model. Indeed the theoretical spatial correlation function $\eqref{coreff}$ is strictly greater than zero so that the dispersion index is expected to grow monotonically. The presence of negative values in the experimental correlation function is explained by the finite diameter of particles. Experiments B took place in a one-dimensional channel whose width equals particle diameters. Thus, there is less probability of finding two particles separated by a distance smaller than the particle diameter, resulting in negative correlation at the diameter scale.

Unfortunately, in none of the experiments presented, the limiting value of the dispersion index $I(\infty)$ is reached for the maximum measurement length. Even experiments J ---designed specially to achieve this purpose--- are unable to reach the final plateau. It is still possible to extrapolate the theoretical expression $\eqref{varmean}$ to predict the behavior of fluctuations at larger scales and to obtain  $I(\infty)$. 

Experimental $K-$functions are displayed in Fig.~\ref{fig:kfunc}(b--f). In Appendix~\ref{app:disp}, we precise how the $K$-function can be estimated experimentally. In all experiments, $K(x)$ is greater than $x$, suggesting that correlations exist between particle locations.

\begin{figure*}
\centerline{\includegraphics[width=17cm]{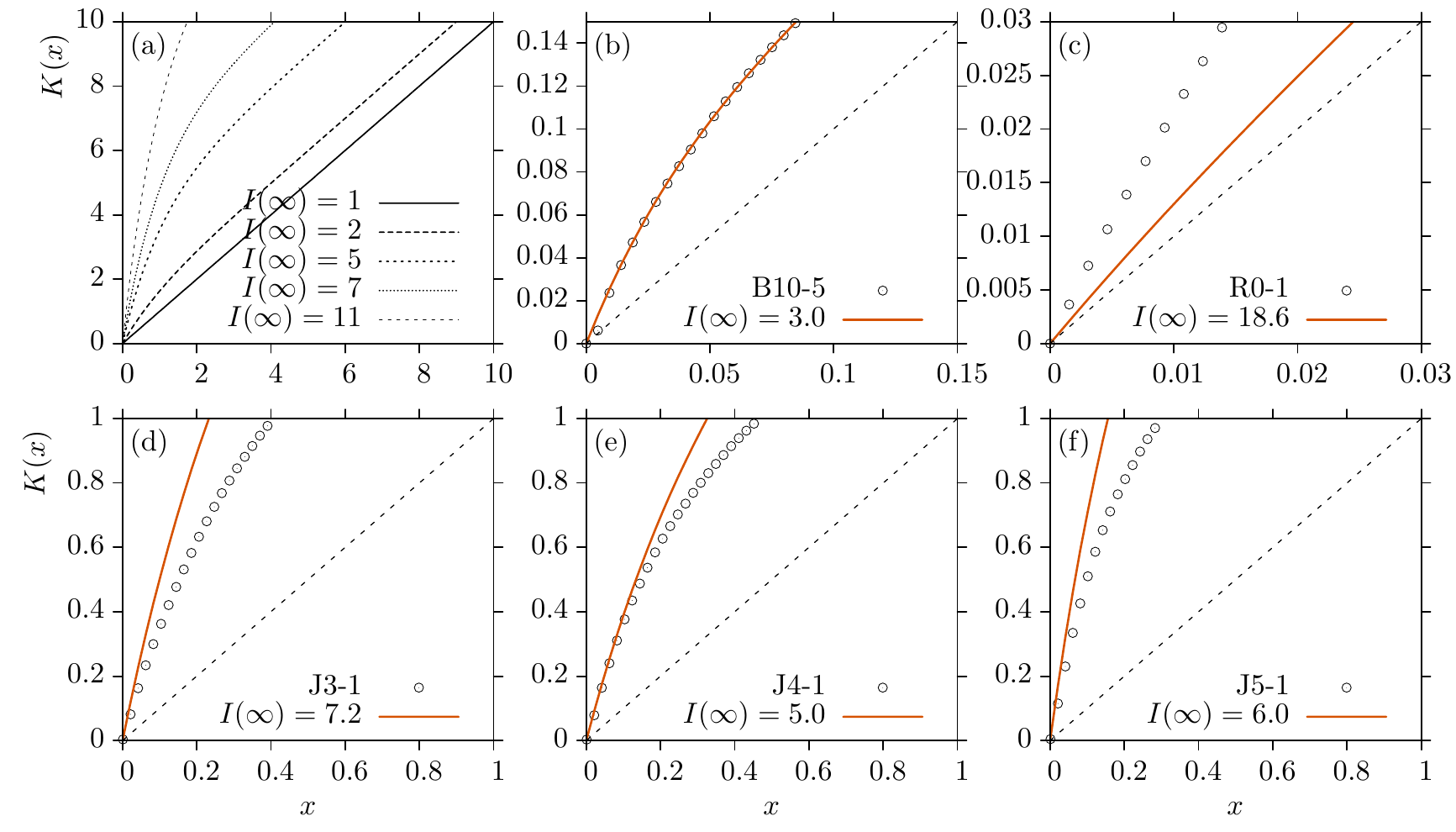}}
\caption{$K$-function. (a) Eq.~$\eqref{Kfunc}$ for $\left\langle{\gamma}\right\rangle_s=1$ and $\ell_c=1$, (b) experiment B, (c) experiment R, (d-e-f) experiments J. The dashed line corresponds to the Poissonian case ($K(x)=x$). Note: fits were calculated from the dispersion index results.}
\label{fig:kfunc}
\end{figure*}

\subsection{Parameter estimates}
{In the following, we show how the parameters  $\bar{u}_s$, $D$, $\lambda$, $\sigma$ and $\mu$ can be estimated given the experimental particle trajectories.}

\begin{figure}
\centerline{\includegraphics[width=8cm]{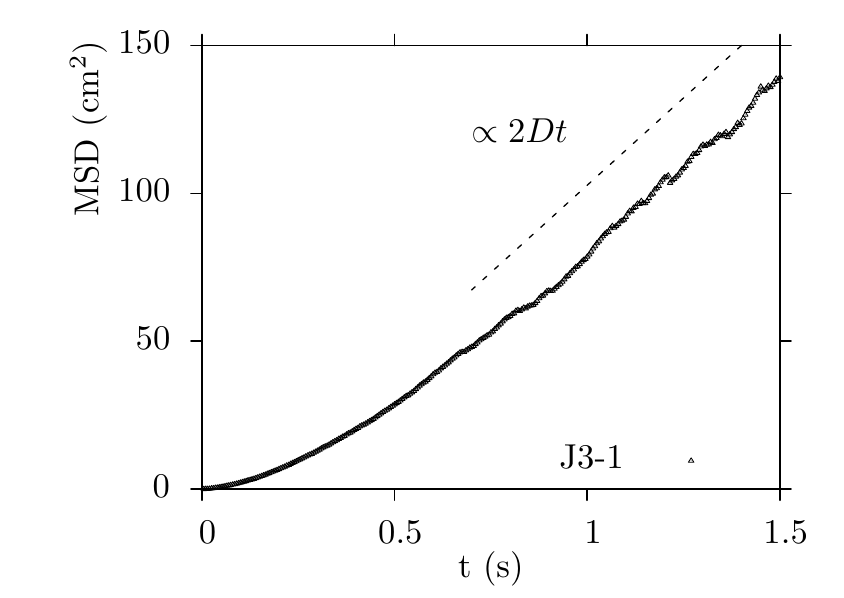}}
\caption{Mean squared displacement (MSD) of particles for experiment J3-1. The dashed line stands for the linear diffusion limit.}
\label{fig:diff}
\end{figure}

{While the average particle velocity $\bar{u}_s$ is simply determined by the arithmetic mean of all instantaneous particle velocities, the diffusivity $D$ can be obtained by computing the experimental mean squared displacement of moving particles. Indeed, the mean squared displacement of particles obeying pure diffusion increases linearly with time: $\left\langle x(t)-\left\langle x(t)  \right\rangle  \right\rangle ^2 = 2D t$ \citep{taylor1922}. As particle motions are correlated, the mean squared displacement is not linear at small time, but tends to the linear diffusion limit at large time \citep{Uhlenbeck1930}. Thus, the limiting diffusivity of particles is obtained by fitting a line to the asymptotic mean squared displacement of particles (Fig.~\ref{fig:diff}). Note that, the mean squared displacement is only computed with the parts of trajectories where the particle moves, excluding the rest periods.}

{The mean deposition rate of particles $\sigma$ is readily obtained by counting deposition events in a given observation window during a given time, then dividing this number by the observation time and by the mean number of moving particles in the window.}

{$\mu$ and $\lambda$ are the only parameters which cannot be estimated independently. The theoretical expressions of first and second moments have to be used to determine their respective values. First, we adjust $\mu$ so that Eq.~$\eqref{varmean}$ matches the experimental dispersion index. Then, $\lambda$ is simply obtained using equation $\eqref{eq:average}$. It is worth highlighting that only two parameters ($\lambda$ and $\mu$) are tuned to fit first and second moments, while the three others are estimated independently. Eq.~$\eqref{varmean}$ captures extremely well the experimental dispersion indices at all scales (Fig.~\ref{fig:disp}(b--f)). Estimated parameters are reported in Table~\ref{tab:table1}. }

{Based on these estimates, we compare the theoretical $K$-function $\eqref{Kfunc}$ to each experiment. From Fig.~\ref{fig:kfunc}(b--f), we can see that the theoretical $K$-function describes experimental data less accurately than the dispersion index does. This could be explained by the fact that an additional parameter, the mean particle activity $\left\langle{\gamma}\right\rangle_s$, is required in Eq.~$\eqref{Kfunc}$. Thus, if the measure of $\left\langle{\gamma}\right\rangle_s$ is biased (as it can be the case for short trajectory samples or for non-homogeneous transport conditions), the agreement between theory and experiments is also biased. Another explanation may be that a two-dimensional $K-$function would be more appropriate for two-dimensional experiments (such as experiment R).}

\subsection{Characteristic lengths and local P\'eclet number}
{Once $D$, $\sigma$ and $\mu$ have been determined, $\ell_c$ can be calculated. Looking at equations $\eqref{coreff}$ and $\eqref{varmean}$, $\ell_c$ defines the scale at which approximately $36.8$\% of the maximum fluctuations above the mean are observed. In other words, to observe at least $95$\% of the total fluctuations, lengths of the order of $20\ell_c$ should be observed experimentally, that is, about $1$~m for experiments B and R, and $3$~m for experiments J. In addition to the technological challenge such a long acquisition length involves, the experimental flume might need to be even longer to avoid effects from the input and output boundary conditions.}

{The saturation length $\ell_{sat}$ (see Eq.~$\eqref{eq:satPe}$) is a good estimate of the length needed for perturbations induced at the boundary to dissipate. It ranges from tens of centimeters in experiments B and R to several meters for experiments J (Table~\ref{tab:table1}).  Unfortunately, no experimental data about saturation length in bed load transport under water are presently available, so that no comparison could be made with these predictions. In any case, it is clear that the boundary conditions of experiments B and J, carried out in relatively short flumes (between $0.5$ and $10\ell_{sat}$), may have an influence on the results.}

{For experimental lengths of the order of $20\ell_c$, it is often impossible to insure the spatial homogeneity of sediment transport. Indeed, the instability of the bed-water interface leads to the development of bedforms of various wavelengths (from centimeters to hundreds of meters) and thus precludes the use of constant parameters.}

{The local P\'eclet number can be calculated with the obtained values of parameters (Table~\ref{tab:table1}). ${\mbox{Pe}}$ is observed to range from 4 (experiment B) to 14 (experiments R and J), showing the variety of modes of transport of bed load. Bed load occurring in experiment B is still strongly diffusive at the correlation length scale while in experiments R and J, it is mostly advective at this scale. }
%
%
%

\section{Summary and discussion}

In this paper, we studied the spatial fluctuations of the number of moving particles per unit bed length, also called the particle activity \citep{Furbish2012(1)}.  These fluctuations have been shown to have a great deal of effect on the measurements of bed load transport rates in both field and experimental surveys \citep{Gomez1990,Dinehart1992,Garcia2000,Cudden2003,Bunte2005}. The model recently proposed by \citet{Ancey2013}, generalizing the probabilistic model of \citet{ANCEY2008} to a spatial dimension, offers a simple theoretical framework to understand and quantify these fluctuations.

{The stochastic model is based on five parameters, most of which have a physical meaning: the entrainment rate $\lambda$, the collective entrainment rate $\mu$, the deposition rate $\sigma$, the average particle velocity $\bar{u}_s$ and the particle diffusivity $D$. The first two have to be estimated via the method of moments (with Eqs.~$\eqref{eq:average}$ and $\eqref{vareq}$) while the three lasts can be calibrated independently. Two characteristic lengths emerge from the model: a saturation length $\ell_{sat}$ (Eq.~$\eqref{eq:satPe}$) quantifying the length needed for particle activity to recover its average equilibrium value, and a correlation length $\ell_c=\sqrt{D/(\sigma-\mu)}$ which describe the typical size of fluctuations in particle activity. We also defined a local P\'eclet number ${\mbox{Pe}}=\bar{u}_s l_c/D$ that describes the relative importance of advection against diffusion of particles at the correlation length. This number plays an important role in the value of the saturation length. }

The stochastic model was tested against various experimental data of particle trajectories, and showed good overall agreement, notably in the description of the dispersion index. 

{This study also provides interesting guidelines for researchers studying the fluctuations of bed load transport rates. To capture 95\% of the fluctuations of particle activity, an experiment should be designed such that it provides a measurement window larger than $20\ell_c$.  The difficulty lies in the fact that $\ell_c$ is not known \textit{a priori}, but it has to be computed after parameters estimation or measured directly if the limiting value of the dispersion index could be reached experimentally.}

{Another issue arises since, at lengths of the order of $20\ell_c$, it is generally difficult to ensure that bed load transport is homogeneous. Indeed, bedforms inexorably develop and migrate, modifying locally the flow and sediment transport. Model parameters may thus vary in time and space, precluding the use of the preceding results, derived in the case of stationary and homogeneous transport conditions.}

{More generally, this leads us to question the use of average equations (such as Eq.~$\eqref{lang2av}$) to describe bed load transport. As fluctuations were shown to span over scales often larger than the ones at which bed load can be considered stationary and homogeneous ---or even at scales larger than the experiment size--- average equations may fail at describing the non-linear interactions that may exist between the fluctuations in particle activity and the changes in bed elevation and water velocity for instance. Consequently, a correct description of bed load transport cannot avoid the modelling of local fluctuations and their interactions with the system boundaries. }

{Imagine now that the stochastic sediment transport equation $\eqref{lang2}$ is coupled with Exner and Saint-Venant or Navier-Stokes equations. By a simultaneous and local description of the bed load activity fluctuations, as well as the fluid flow and the bed surface evolution, we may be able to give a more accurate picture of the whole transport process. }

{The proposed model may also be generalized to a second spatial dimension. Indeed, the motion of particles is only rarely unidirectional, as particle collisions and turbulent flow drag tend to modify particle trajectories \citep{Seizilles2014}. Thus, a dispersion of particles in the direction normal to the mean sediment velocity vector may also occur. The corresponding cross-stream diffusivity is expected to be different than the streamwise diffusivity, so that the overall diffusion process might be anisotropic. Still, owing to the overall linearity of the equations, the addition of a cross-stream diffusion term is straightforward.}

 \begin{notation}
 $\left\langle \bullet \right\rangle$ & Ensemble average. \\
 $\left\langle  \bullet \right\rangle_s$ & Ensemble average in steady state and homogeneous conditions. \\
 $\left\langle \bullet,\bullet \right\rangle$ & Covariance of two random variables.  For instance, $\left\langle X,X \right\rangle=\left\langle X^2\right\rangle - \left\langle X\right\rangle^2$.\\
$N_i/n_i$       &  Random variable/number of moving particles in cell $i$.\\
$\Delta x$      &  Cell length in m.\\
$\gamma(x,t)$    &  Density of moving particles at location $x$ and time $t$ in particles~m$^{-1}$.\\
$\boldsymbol{n}$		&  Vector of the number of moving particles in each cell.\\
$x_i$			&  Position of the center of the cell $i$. \\
$\lambda$		&  Average particle entrainment rate per meter length in particles~m$^{-1}$~s$^{-1}$.\\
$\mu$     		&  Average collective entrainment rate in s$^{-1}$.\\
$\sigma$  		&  Average particle deposition rate in s$^{-1}$.\\
$D$      		&  Macroscopic diffusivity in m$^2$~s$^{-1}$.\\
$\bar{u}_s$       		&  Mean particle velocity in m~s$^{-1}$.\\
$d$      		&  Local diffusivity in s$^{-1}$.\\
$\mathbf{r}_i^\pm$ & Vector whose all but one value are zero: $r_i=\pm 1$, $r_k=0$ for $k\neq i$.\\
$a_i$ 			& Poisson rate in cell $i$ in the Poisson representation of $n_i$.\\
$\boldsymbol{a}$  & Vector of Poisson rates in each cell.\\
$f$	& Pseudo-density function of $\boldsymbol{a}$. \\
$\mbox{d}  W_i(t)$	& derivative of a temporal Wiener process (white noise). \\
$\eta(x,t)$      &  Poisson density of moving particles at location $x$ and time $t$ in particles~m$^{-1}$.\\
$\mbox{d}  W(x,t)$	& derivative of a spatio-temporal Wiener process (two-dimensional white noise). \\
$\left\langle{\gamma}\right\rangle_s$      	 &  Steady-state homogeneous average density of moving particles in particles~m$^{-1}$.\\
$\left\langle{\eta}\right\rangle_s$      	 &  Steady-state homogeneous average Poisson density of moving particles in particles~m$^{-1}$.\\
$\ell_c$        	 & Correlation length in m.\\
$\ell_{sat}$        	 & Saturation length in m.\\
${\mbox{Pe}}$      	 & Local P\'eclet number (dimensionless number).\\
$I$				 & Index of dispersion. \\
$L$				 & Length of the observation window in m. \\
$\tilde{L}$	 & dimensionless length ($L/\ell_c$) of the observation window. \\
$K$  			 & $K$-function.
\end{notation}

\begin{acknowledgments}
This work was supported by the Swiss National Science Foundation under grant number 200021\_129538 (a project called ``The Stochastic Torrent: stochastic model for bed load transport on steep slope''), the R'Equip grant number 206021\_133831 and by the competence center in Mobile Information and Communication Systems (grant number 5005-67322, MICS project).
\end{acknowledgments}

\newpage
\appendix
\section{Link to the BCRE model}\label{app:BCRE}
The BCRE model presented  by \citet{bouchaud1995} gives the density of rolling grains $\mathcal{R}$ as the solution of:
\begin{equation}
\partial_t \mathcal{R} + \nabla\left( V \mathcal{R}\right) = \nabla^2\left (D \mathcal{R}\right) -  \mathcal{R} \alpha \nabla h,
\label{eq:BCRE}
\end{equation} 

where $\nabla h$ stands for the bed slope variations close to the angle of repose and $\alpha$ is a constant. Thus, in their model, when the slope is bigger than the angle of repose ($\nabla h < 0$) the second term on the left-hand side acts as a source in the equation. In that case, the number of rolling grains increases exponentially, leading to a local avalanche. On the contrary, when the slope is less than the angle of repose ($\nabla h >0$), grains are mainly deposited, causing the avalanche to stop ($\mathcal{R}=0$). The resemblance with Eq.~$\eqref{lang2}$ is striking. In the latter, an exponential increase in the number moving particles occurs when the collective entrainment rate is greater than or equal to the deposition rate ($\mu \geq \sigma$). In contrast to $\eqref{eq:BCRE}$, when deposition is greater than collective entrainment, a non trivial steady-state solution exists, due to the uncorrelated particle entrainment process (with rate $\lambda$). 

Our model could thus be seen as a ``BCRE'' model that includes an additional random perturbation. Though the present work concerns bed load transport, and we restrict ourselves to the steady-state case ($\mu<\sigma$), the limit $\mu\rightarrow\sigma$ might be of particular interest for other granular flows. In particular, we suggest that $\eqref{lang2}$ may also be applicable to certain dry and dilute granular flow, and thus may allow for their statistical description.
\section{Link to point processes}\label{app:PP}
It is possible to draw an analogy between equation $\eqref{lang2}$, obtained in the framework of birth-death Markov processes through Poisson representation, and the point process framework. Indeed, point processes are often defined by their rate function $\eta(x,t)$ \citep{coxbook}. The simplest case is when the rate function is constant in time and space, resulting in a Poisson point process. When the rate function is a function of space and/or time, the process is called an inhomogeneous Poisson point process. Eventually, when the rate function is also a random variable, the process is called a doubly stochastic process, or Cox process \citep{coxbook}.  This is the case with Eq.~$\eqref{lang2}$. To summarize, starting from a multivariate Markov process defined on lattice cells and described by a master equation, we end up with a model belonging to a general class of point processes, called doubly stochastic processes. 
 
\begin{figure*}
\centerline{\includegraphics[width=17cm]{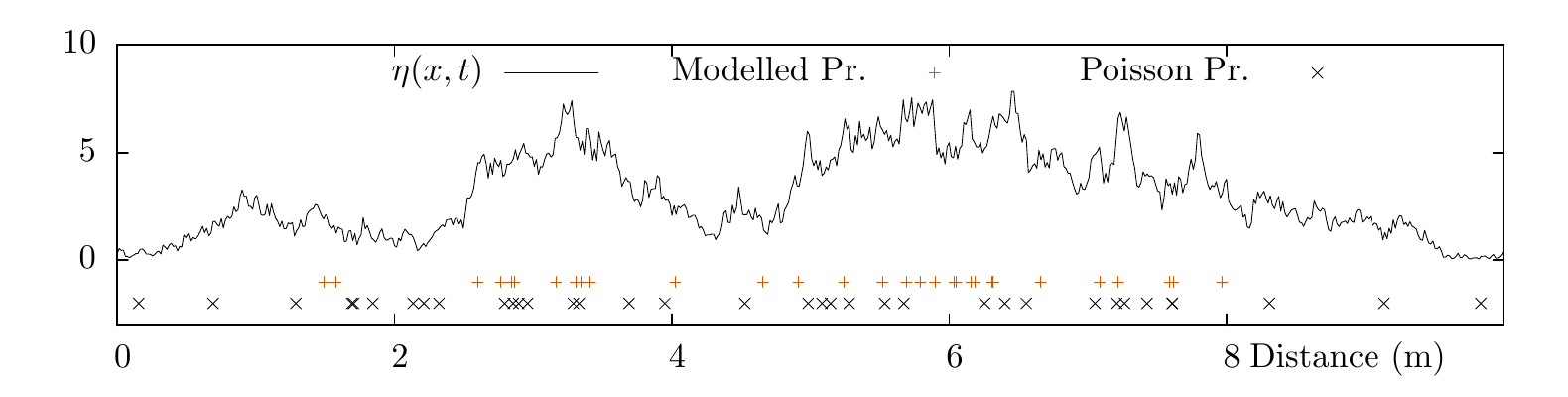}}
\caption{Example simulation of the rate process (Eq.~$\eqref{lang2}$) and corresponding possible realization of particle positions. We also plot the Poissonian case (with the same mean rate) for comparison. Model parameters are $\lambda=0.05$~particles~m$^{-1}$, $\mu=9.99$~s$^{-1}$, $\sigma=10$~s$^{-1}$, $\bar{u}_s=0.1$~m~s$^{-1}$ and $D=0.008$~m$^2$~s$^{-1}$}
\label{sim}
\end{figure*}

We now show how it is possible to simulate a probable realization of particle positions from Eq.~$\eqref{lang2}$. As noted earlier, by means of the Poisson representation, $\eta(x,t)$ can be interpreted as the random rate of a Poisson distribution. First, we need to compute Eq.~$\eqref{lang2}$, to get a realization of $\eta(x)$ at a given time $t$. This can be achieved using standard methods for stochastic differential equations {(for instance an Euler--Maruyama scheme, which is an explicit finite-difference numerical scheme for stochastic equations \citep{kloedenBook})}. Once we get a realization of $\eta(x)$, we proceed as follows. We choose a constant $C>\eta(x)$ and compute a realization of point positions according to a Poisson process with rate $C$. {This can be achieved by drawing $C L$ random point locations from the uniform distribution between $0$ and $L$, $L$ being the length of the computation area.} Then, each point is randomly selected or discarded according to the criteria: 

\begin{eqnarray}
\mbox{if }r&<&\eta(x_k)/C, \mbox{ keep point}; \nonumber\\ 
\mbox{if }r&>&\eta(x_k)/C, \mbox{ delete point};\nonumber
\end{eqnarray} 

where $r$ is drawn from a uniform distribution in $[0,1]$. The remaining points form a possible observation of particle positions according to the model (Fig.~\ref{sim}).

In Fig.~\ref{sim}, it is possible to observe the clustering of particles around the region of high $\eta(x)$ values, while for the Poisson process, particles positions are purely random so no clustering appears. The clustering of particles is a special feature of our model (when $\mu>0$) and can be quantified by the study of the second moment. 

\section{Theoretical developments}
\subsection{Spatial correlation function}\label{app:space}
Let $g(x,x',t)$ denote the spatial correlation function of the Poisson density variable $\eta(x,t)$. By definition we have

\begin{eqnarray}
g(x,x',t)&=&\left\langle{\eta(x,t),\eta(x',t)}\right\rangle \nonumber \\
&=&\left\langle{\eta(x,t)\eta(x',t)}\right\rangle -\left\langle{\eta(x,t)}\right\rangle \left\langle{\eta(x',t)}\right\rangle
\label{defig}.
\end{eqnarray} 
Taking the differential of $g$ and using It\=o's calculus rules (an equivalent of the chain rule for stochastic equations),

\begin{eqnarray}
\mbox{d}  g(x,x',t)&=& \mbox{d}  \left\langle{\eta(x,t)\eta(x',t)}\right\rangle \nonumber \\
&=& \left\langle{ \mbox{d} \eta(x,t)\eta(x',t)}\right\rangle+ \left\langle{\eta(x,t) \mbox{d} \eta(x',t)}\right\rangle \\
& +& \left\langle{\mbox{d}  \eta(x,t) \mbox{d} \eta(x',t)}\right\rangle.  \nonumber 
\end{eqnarray} 
Note that $\mbox{d}  \left(\left\langle{\eta(x,t)}\right\rangle \left\langle{\eta(x',t)}\right\rangle\right)$ is zero by definition of the average. It comes

\begin{eqnarray}
&&\mbox{d} g(x,x',t)= D \left(\partial^2/\partial x^2+  \partial^2/\partial x'^2\right) \left\langle{\eta(x,t)\eta(x',t)}\right\rangle \mbox{d}t\nonumber \\
&&-\bar{u}_s  \left(\partial/\partial x + \partial/\partial x'\right) \left\langle{\eta(x,t)\eta(x',t)}\right\rangle \mbox{d}t  \\
&&-2(\sigma-\mu)\left\langle{\eta(x,t)\eta(x',t)}\right\rangle \mbox{d}t + 2\left\langle{\eta}\right\rangle_s\left(\lambda+\mu \delta(x-x')\right) \mbox{d}  t. \nonumber
\label{geq}
\end{eqnarray} 
In a spatially homogeneous situation, $ g(x,x',t)$ is a function of $r=|x-x'|$ only, which we call $g(r,t)$. Thus, substituting Eq.~$\eqref{defig}$ into Eq.~$\eqref{geq}$, we obtain
\begin{equation}
\frac{1}{2}\frac{\partial g(r,t)}{\partial t}=D \frac{\partial^2 g(r,t)}{\partial r^2}-(\sigma-\mu)g(r,t)+ \mu\left\langle{\gamma}\right\rangle_s \delta(r).
\label{geqhom}
\end{equation} 
The advection term disappears because $\partial/\partial x=-\partial/\partial x'$. Thus, the spatial correlation has no dependence on the mean velocity of particles.
We look for the stationary behavior of $\eqref{geqhom}$, which is
\begin{equation}
D \frac{\partial^2 g_s(r)}{\partial r^2}-(\sigma-\mu)g_s(r)+ \mu\left\langle{\gamma}\right\rangle_s \delta(r)=0,
\label{appeqspat}
\end{equation} 
with $r=|x-x'|$. We can simplify Eq.~$\eqref{appeqspat}$ by rescaling the variable $r$ by $ \tilde{r}=r/\ell_c$ where $\ell_c=\sqrt{D/(\sigma-\mu)}$. It yields 
\begin{equation}
\frac{\partial^2 g_s( \tilde{r})}{\partial \tilde{r}^2}-g_s( \tilde{r})+\frac{\left\langle{\gamma}\right\rangle_s}{\ell_c} \frac{\mu}{\sigma-\mu} \delta( \tilde{r})=0.
\label{appeqspat2}
\end{equation} 
By means of Fourier transform, we obtain the algebraic equation
\begin{equation}
G(\omega)=\frac{\left\langle{\gamma}\right\rangle_s}{\ell_c} \frac{\mu }{\sigma-\mu} \frac{1}{\omega^2+1},
\label{four}
\end{equation} 
where $G(\omega)$ is the Fourier transform of $g_s(\tilde{r})$.
The Fourier inverse of Eq.~$\eqref{four}$, is given by
\begin{equation}
g_s( \tilde{r})= \frac{ \left\langle{\gamma}\right\rangle_s}{2 \ell_c} \frac{\mu }{\sigma-\mu}\exp{\left(-| \tilde{r}|\right)}.
\end{equation} 
Hence
\begin{equation}
\left\langle \eta(x),\eta(x')\right\rangle_s=  \frac{\left\langle{\gamma}\right\rangle_s}{2 \ell_c} \frac{\mu }{\sigma-\mu}\exp{\left(-\frac{|x-x'|}{\ell_c}\right)}.
\end{equation} 
Second moment of $\eta$ and $\gamma$ are connected through the simple relationship \citep{Gardiner1977}
\begin{equation}
\left\langle{\eta(x,t),\eta(x',t)}\right\rangle=\left\langle \gamma(x,t),\gamma(x',t)\right\rangle-\delta(x-x')\left\langle{\gamma(x,t)}\right\rangle, \nonumber
\label{relpoiss}
\end{equation} 
so that Eq.~$\eqref{coreff}$ is recovered.

\subsection{Spatial fluctuations}\label{A4}
We wish to compute the integral
\begin{equation}
\mbox{Var}[N(L)]= \int_L \int_L \left\langle\gamma(x,t),\gamma(x',t)\right\rangle_s \mbox{d}x~ \mbox{d}x'.\nonumber
\label{vareqA}
\end{equation} 

That is

\begin{eqnarray}
\mbox{Var}[N(L)]&=&\left\langle{\gamma}\right\rangle_s L \nonumber \\
&+&\frac{\left\langle{\gamma}\right\rangle_s}{2 \ell_c}\frac{\mu}{\sigma-\mu} \int_{-L/2}^{L/2} \int_{-L/2}^{L/2} \mbox{e}^{-|x-x'|/\ell_c} \mbox{d}x~ \mbox{d}x'. \nonumber
\label{vareqA1}
\end{eqnarray} 

The value of the integral can be obtained by using

\begin{eqnarray}
&&\int_{-L/2}^{L/2} \int_{-L/2}^{L/2} \mbox{e}^{|x-x'|/\ell_c} \mbox{d}x~ \mbox{d}x' = \nonumber\\
&&\int_{-L/2}^{L/2} \left[ \int_{-L/2}^{x} \mbox{e}^{(x-x')/\ell_c} \mbox{d}x+\int_{x}^{L/2} \mbox{e}^{-(x-x')/\ell_c} \mbox{d}x\right] \mbox{d}x' = \nonumber\\
&&\ell_c \int_{-L/2}^{L/2} \left[ 2-\mbox{e}^{L/(2\ell_c)}\left(\mbox{e}^{-x'/\ell_c}+\mbox{e}^{x'/\ell_c}\right)\right] \mbox{d}x' = \nonumber\\
&& 2 \ell_c^2\left(L/\ell_c+\mbox{e}^{-L/\ell_c}-1\right). \nonumber
\label{vareqA2}
\end{eqnarray} 

Thus
\begin{equation}
\mbox{Var}[N(L)]= \left\langle{\gamma}\right\rangle_s L+\left\langle{\gamma}\right\rangle_s \ell_c \frac{\mu}{\sigma-\mu}\left(L/\ell_c+\mbox{e}^{-L/\ell_c}-1\right). \nonumber
\label{vareqA3}
\end{equation}

\newpage
\section{Experiments}\label{app:setups}
\subsection{Experiments B}
These experiments were carried out in a narrow steep flume where sediment consisted of glass beads of equal size (6 mm). Particle transport was completely two-dimensional; this allowed \citet{ANCEY2004} to take pictures through the side wall and detect and track individual particles via image processing.
Camera resolution was $640 \times 192$ pixels with a frame rate of 129.2~frame per seconds (fps). Each sequence comprised 8000 images corresponding to a duration of approximately 1 min. The acquisition length was 22.5~cm, for a resolution of 0.3~mm/pixel. Thus this imaging technique covers about 2 orders of magnitude in space.
For further  information on the experimental conditions, the reader is referred to \citep{ANCEY2004,ANCEY2008}. 
 
\begin{figure}
 \centerline{\includegraphics[width=8cm]{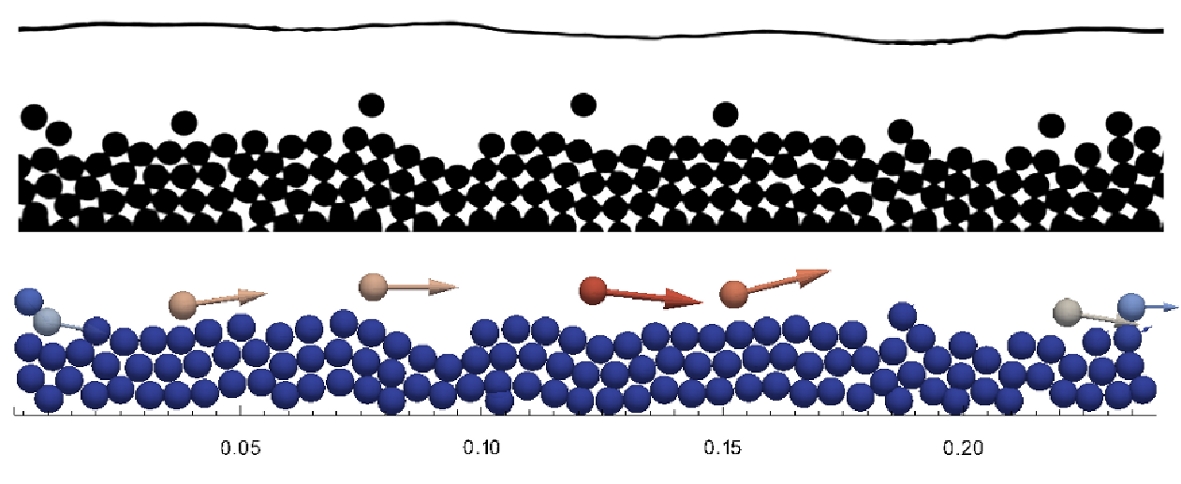}}
 \caption{Image recorded during a B experiment (top) and visualization of particles velocity vectors after image processing (bottom). Units are in meter.}
 \label{Tobias3d}
 \end{figure}

Fig.~\ref{Tobias3d} shows an example of a recorded image and the corresponding reconstruction of particle positions and velocities using image processing.

\subsection{Experiment R}
\citet{Furbish2012(2)} presented a set of experiments where particle trajectories were sampled in a two-dimensional window of the bed viewed from the top. High-speed imaging at 250~fps over a 7.57~cm (streamwise) by 6.05~cm (cross-stream) bed-surface domain, and with 1280$\times$1024 pixels resolution provided the basis for tracking particle motions (with a precision of 0.06~mm/pixel). Bed material consisted of relatively uniform coarse sand with an average diameter of $d_{50}=0.5$~mm.

\begin{figure}
 \centerline{\includegraphics[width=8cm]{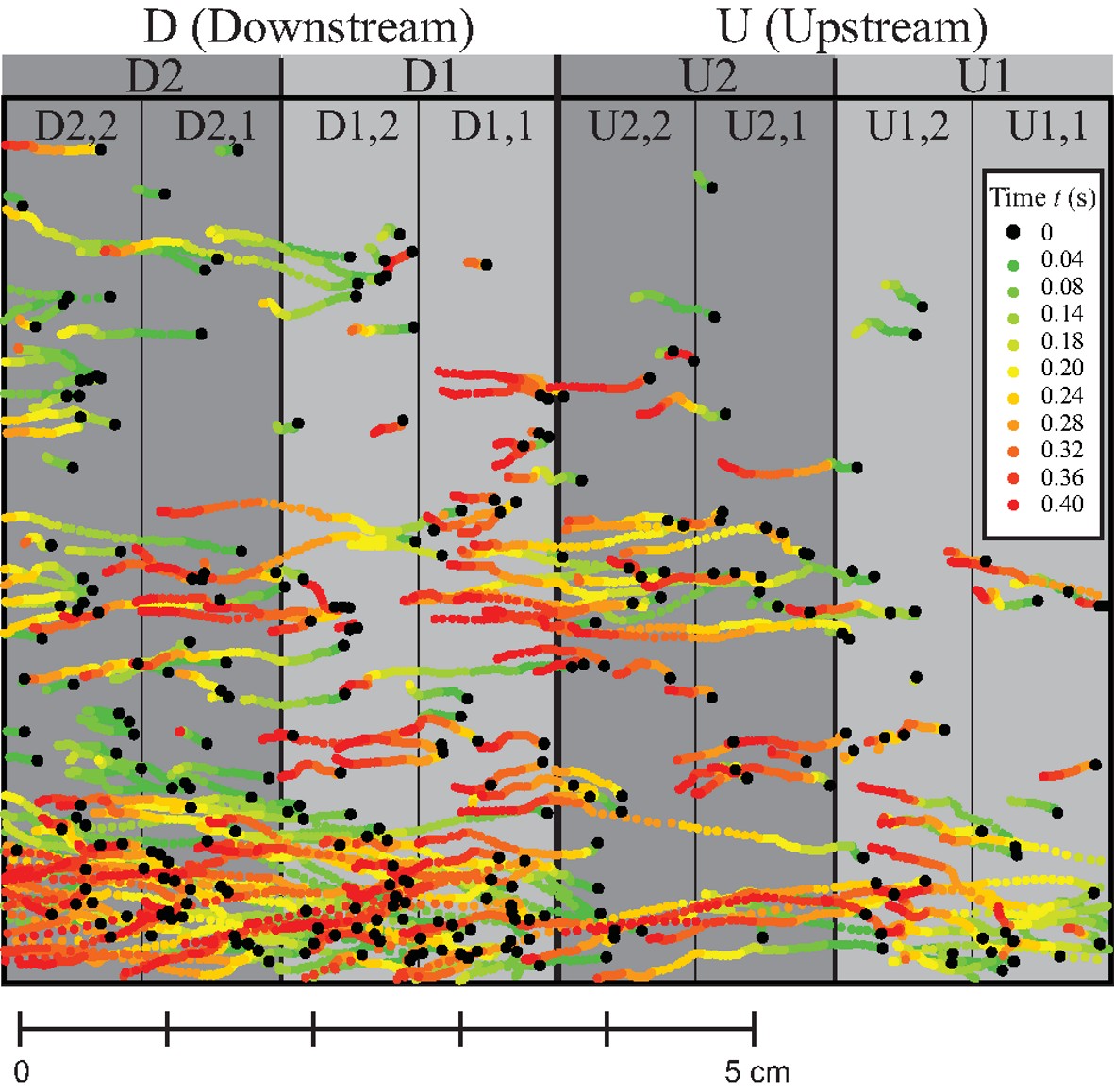}}
 \caption{Map view of experiments R showing particle motions occurring during the 0.4 sec time series; note the clustering of motions, partly reflecting effects of the turbulent sweeps. (Reproduced from \citet{Furbish2012(2)} with the authorization of the authors and AGU.)}
 \label{Furbish}
 \end{figure} 

The data set involved one experiment with a total duration of 0.4 seconds, i.e. 100 frames (Fig. \ref{Furbish}). 
In contrast to the three other data sets, experiment R concerns relatively small particles (sand) over mild slope (the slope is not given in \citep{Furbish2012(2)} but the Froude number is much lower than unity).

\subsection{Experiments J}
 
The originality of this data set compared to the two others lies in its high temporal and spatial resolutions. Two cameras of $1280\times 200$ pixels resolution, placed side by side, took pictures from the transparent side wall at a rate of 200 fps. 
The length of the observation window was slightly less than 1~m (with a precision of about 0.4~mm/pixel) while the duration of a sequence was 150 seconds (30 000 images). For each experiment, four film sequences were repetitively taken to insure good statistical results.

Experiments were carried out in a 2.5-m-long flume. The erodible bed was made of natural sediment particles with mean diameter of 8~mm. The flume was 3.5~cm wide and the water depth was ranging from 3 to 4~cm during experiments. The channel slope ranged between 3\% and 5\%. The flow was fully supercritical. Small antidunes were occasionally growing and propagated upstream, but the bed remained nearly flat in all experiments. As the channel width  to flow depth ratio was relatively small ($B/h\sim1$), the fraction of the shear stress taken by the bed was certainly reduced, because of the increased side wall friction. Experimental studies \citep{Knight1981} report a drop of about 40 to 60\% of the bed shear stress for such aspect ratio. It is thus hard to determine precisely the experimental Shield stress without any direct flow velocity measurements.

Image processing and automatic particle tracking were then performed on these images (Fig.~\ref{newexp}).  The processing steps from raw images to particle trajectories were the following:

\begin{enumerate}
\item First the raw images were processed using the powerful yet simple method of median background subtraction \citep{Yilmaz2006,radice2006}. This allows a distinction between an immobile background (made up of particle resting on the bed) and a moving foreground (the moving particles).
\item An algorithm was then used to detect the centroid position of the moving particles in the foreground images. This was achieved after thresholding the foreground image and computing properties of connected regions (such as area, barycentre, eccentricity...)
\item Moving particles between two consecutive images were then associated into trajectories. The Hungarian algorithm was used here to obtain the best combinations. In case of conflict (for instance, if two particles are assigned to the same particle in the following frame), the trajectory was supposed to end and a new trajectory was built.
\item Finally, to reconstruct broken trajectories, a Kalman filter was applied to each missing measurements and overlapping trajectories were merged.
\end{enumerate}

\subsection{Experimental dispersion index and $K$-function}\label{app:disp}

The experimental dispersion index at length $L$ is estimated as follows.  First, we randomly choose $p$ sub-windows of equal length $L$ inside the whole available observation region. As the sub-windows are selected randomly, they possibly overlap. Then, we construct the vector $N_{i,k}$ of the number of moving particles found in the sub-window $i$ ($i=1,\cdots ,p$) at frame $k$ ($k=1,\cdots ,f$, $f$ being the total number of video frames). Var[$N(L)$] and Mean[$N(L)$] are then estimated with all $N_{i,k}$ samples. Typically, we chose $p=20$. In experiments J, the dispersion index is thus estimated over more than 2 millions of samples. Note that, such a large number of samples is necessary to get unbiased estimates of moments, since samples may not be independent of each other. Similarly, the dispersion index of experiment R is estimated over 2000 samples and the dispersion index of experiment B is estimated over 640~000 samples.

The procedure to compute the experimental $K$-function is given in \citet{RipleyK}. In the latter paper, several methods are presented to prevent the problems arising at the boundaries of the image. Indeed, for a particle located close to the image boundary, the number of particles found in a circle of radius $x$ larger than the particle-boundary distance may be underestimated. To prevent this, we chose to limit the computation of $K(x)$ to particles located at a minimum distance $x$ from the image boundary.

\newpage
\end{article}
%

\begin{thebibliography}{48}
\providecommand{\natexlab}[1]{#1}
\expandafter\ifx\csname urlstyle\endcsname\relax
  \providecommand{\doi}[1]{doi:\discretionary{}{}{}#1}\else
  \providecommand{\doi}{doi:\discretionary{}{}{}\begingroup
  \urlstyle{rm}\Url}\fi

\bibitem[{\textit{Ancey}(2010)}]{ANCEY2010}
Ancey, C. (2010), Stochastic modeling in sediment dynamics: Exner equation for
  planar bed incipient bed load transport conditions, \textit{J. Geophys.
  Res.}, \textit{115}, F00A11, \doi{10.1029/2009JF001260}.

\bibitem[{\textit{Ancey and Heyman}(2014)}]{Ancey2013}
Ancey, C., and J.~Heyman (2014), A microstructural approach to bed load
  transport: mean behaviour and fluctuations of particle transport rates,
  \textit{J. Fluid Mech.}, \textit{744}, 129--168, \doi{10.1017/jfm.2014.74}.

\bibitem[{\textit{Ancey et~al.}(2006)\textit{Ancey, B\"ohm, Jodeau, and
  Frey}}]{Ancey2006}
Ancey, C., T.~B\"ohm, M.~Jodeau, and P.~Frey (2006), Statistical description of
  sediment transport experiments, \textit{Phys. Rev. E}, \textit{74}(1),
  011,302, \doi{10.1103/PhysRevE.74.011302}.

\bibitem[{\textit{Ancey et~al.}(2008)\textit{Ancey, Davison, B{\"o}hm, Jodeau,
  and Frey}}]{ANCEY2008}
Ancey, C., A.~C. Davison, T.~B{\"o}hm, M.~Jodeau, and P.~Frey (2008),
  Entrainment and motion of coarse particles in a shallow water stream down a
  steep slope, \textit{J. Fluid Mech.}, \textit{595}, 83--114,
  \doi{10.1017/S0022112007008774}.

\bibitem[{\textit{B{\"o}hm et~al.}(2004)\textit{B{\"o}hm, Ancey, Frey, Reboud,
  and Ducottet}}]{ANCEY2004}
B{\"o}hm, T., C.~Ancey, P.~Frey, J.~Reboud, and C.~Ducottet (2004),
  Fluctuations of the solid discharge of gravity-driven particle flows in a
  turbulent stream, \textit{Phys. Rev. E}, \textit{69}(6), 061,307,
  \doi{10.1103/PhysRevE.69.061307}.

\bibitem[{\textit{Bouchaud et~al.}(1995)\textit{Bouchaud, Cates, Prakash, and
  Edwards}}]{bouchaud1995}
Bouchaud, J.-P., M.~E. Cates, J.~R. Prakash, and S.~F. Edwards (1995),
  Hysteresis and metastability in a continuum sandpile model, \textit{Phys.
  Rev. Lett.}, \textit{74}(11), 1982--1985, \doi{10.1103/PhysRevLett.74.1982}.

\bibitem[{\textit{Bunte and Abt}(2005)}]{Bunte2005}
Bunte, K., and S.~R. Abt (2005), Effect of sampling time on measured gravel bed
  load transport rates in a coarse-bedded stream, \textit{Water Resour. Res.},
  \textit{41}(11), W11,405, \doi{10.1029/2004WR003880}.

\bibitem[{\textit{Celik et~al.}(2010)\textit{Celik, Diplas, Dancey, and
  Valyrakis}}]{celik2010}
Celik, A., P.~Diplas, C.~Dancey, and M.~Valyrakis (2010), Impulse and particle
  dislodgement under turbulent flow conditions, \textit{Phys. Fluids},
  \textit{22}(4), 046,601, \doi{10.1063/1.3385433}.

\bibitem[{\textit{Charru}(2006)}]{charru2006}
Charru, F. (2006), Selection of the ripple length on a granular bed sheared by
  a liquid flow, \textit{Phys. Fluids}, \textit{18}(12), 121,508,
  \doi{10.1063/1.2397005}.

\bibitem[{\textit{Cox and Isham}(1980)}]{coxbook}
Cox, D.~R., and V.~Isham (1980), \textit{Point Processes}, Monograph on applied
  statistics and probability, Chapman \& Hall, London, \doi{0-412-21910-7}.

\bibitem[{\textit{Cudden and Hoey}(2003)}]{Cudden2003}
Cudden, J.~R., and T.~B. Hoey (2003), The causes of bedload pulses in a gravel
  channel: the implications of bedload grain-size distributions, \textit{Earth
  Surf. Processes}, \textit{28}(13), 1411--1428, \doi{10.1002/esp.521}.

\bibitem[{\textit{Detert et~al.}(2010)\textit{Detert, Weitbrecht, and
  Jirka}}]{DETERT}
Detert, M., V.~Weitbrecht, and G.~Jirka (2010), {Laboratory measurements on
  turbulent pressure fluctuations in and above gravel beds}, \textit{J.
  Hydraul. Eng.}, \textit{136}(10), 779--789,
  \doi{10.1061/(ASCE)HY.1943-7900.0000251}.

\bibitem[{\textit{Dinehart}(1992)}]{Dinehart1992}
Dinehart, R.~L. (1992), Evolution of coarse gravel bed forms: Field
  measurements at flood stage, \textit{Water Resour. Res.}, \textit{28}(10),
  2667--2689, \doi{10.1029/92WR01357}.

\bibitem[{\textit{Drake et~al.}(1988)\textit{Drake, Shreve, Dietrich, Whiting,
  and Leopold}}]{Drake1988}
Drake, T.~G., R.~L. Shreve, W.~E. Dietrich, P.~J. Whiting, and L.~B. Leopold
  (1988), Bedload transport of fine gravel observed by motion-picture
  photography, \textit{J. Fluid Mech.}, \textit{192}, 193--217,
  \doi{10.1017/S0022112088001831}.

\bibitem[{\textit{Dwivedi et~al.}(2011)\textit{Dwivedi, Melville, Shamseldin,
  and Guha}}]{dwivedi2011}
Dwivedi, A., B.~Melville, A.~Shamseldin, and T.~Guha (2011), Analysis of
  hydrodynamic lift on a bed sediment particle, \textit{J. Geophys. Res.},
  \textit{116}(F2), F02,015, \doi{10.1029/2009JF001584}.

\bibitem[{\textit{Einstein}(1937)}]{Einstein1937}
Einstein, H.~A. (1937), \textit{Der {G}eschiebetrieb als
  {W}ahrscheinlichkeitproblem ({B}edload transport as a probability problem)},
  English translation by W.W. Sayre, in {Sedimentation}, Ed. H.W. Shen, Fort
  Collins, Colorado, 1972, C1--C105, ETHZ.

\bibitem[{\textit{Einstein}(1950)}]{Einstein1950}
Einstein, H.~A. (1950), \textit{The bed-load function for sediment
  transportation in open channel flows}, Tech. Bul. No. 1026, U.S. Dept. of
  Agriculture, Washington, D. C.

\bibitem[{\textit{Furbish and Schmeeckle}(2013)}]{Furbish2013}
Furbish, D., and M.~Schmeeckle (2013), A probabilistic derivation of the
  exponential-like distribution of bed load particle velocities, \textit{Water
  Resour. Res.}, \textit{49}(3), 1537--1551, \doi{10.1002/wrcr.20074}.

\bibitem[{\textit{Furbish et~al.}(2012)\textit{Furbish, Haff, Roseberry, and
  Schmeeckle}}]{Furbish2012(1)}
Furbish, D., P.~Haff, J.~Roseberry, and M.~Schmeeckle (2012), A probabilistic
  description of the bed load sediment flux: 1. theory, \textit{J. Geophys.
  Res.-earth}, \textit{117}(F3), F03,031, \doi{10.1029/2012JF002352}.

\bibitem[{\textit{Garcia et~al.}(2000)\textit{Garcia, Laronne, and
  Sala}}]{Garcia2000}
Garcia, C., J.~B. Laronne, and M.~Sala (2000), Continuous monitoring of bedload
  flux in a mountain gravel-bed river, \textit{Geomorphology},
  \textit{34}(1–2), 23 -- 31, \doi{10.1016/S0169-555X(99)00128-2}.

\bibitem[{\textit{Gardiner and Chaturvedi}(1977)}]{Gardiner1977}
Gardiner, C., and S.~Chaturvedi (1977), The {P}oisson representation. {I}. a
  new technique for chemical master equations, \textit{J. Stat. Phys.},
  \textit{17}, 429--468, \doi{10.1007/BF01014349}.

\bibitem[{\textit{Gomez et~al.}(1990)\textit{Gomez, Hubbell, and
  Stevens}}]{Gomez1990}
Gomez, B., D.~W. Hubbell, and H.~H. Stevens (1990), At-a-point bed load
  sampling in the presence of dunes, \textit{Water Resour. Res.},
  \textit{26}(11), 2717--2731, \doi{10.1029/WR026i011p02717}.

\bibitem[{\textit{Heyman et~al.}(2013)\textit{Heyman, Mettra, Ma, and
  Ancey}}]{Heyman2013}
Heyman, J., F.~Mettra, H.~B. Ma, and C.~Ancey (2013), Statistics of bedload
  transport over steep slopes: Separation of time scales and collective motion,
  \textit{Geophys. Res. Lett.}, \textit{40(1)}, 128–133,
  \doi{10.1029/2012GL054280}.

\bibitem[{\textit{Hoey}(1992)}]{hoey1992}
Hoey, T. (1992), {Temporal variations in bedload transport rates and sediment
  storage in gravel-bed rivers}, \textit{Prog. Phys. Geog.}, \textit{16}(3),
  319--338, \doi{10.1177/030913339201600303}.

\bibitem[{\textit{Jerolmack and Mohrig}(2005)}]{Jerolmack2005(2)}
Jerolmack, D.~J., and D.~Mohrig (2005), A unified model for subaqueous bed form
  dynamics, \textit{Water Resour. Res.}, \textit{41}(12), W12,421,
  \doi{10.1029/2005WR004329}.

\bibitem[{\textit{Kloeden and Platen}(2011)}]{kloedenBook}
Kloeden, P.~E., and E.~Platen (2011), \textit{Numerical Solution of Stochastic
  Differential Equations}, Springer, New York.

\bibitem[{\textit{Knight}(1981)}]{Knight1981}
Knight, D. (1981), Boundary shear in smooth and rough channels, \textit{J
  Hydraul Div}, \textit{107}(7), 839--851.

\bibitem[{\textit{Kuhnle and Southard}(1988)}]{kuhnle}
Kuhnle, R., and J.~Southard (1988), Bed load transport fluctuations in a gravel
  bed laboratory channel, \textit{Water Resour. Res.}, \textit{24}(2),
  247--260, \doi{10.1029/WR024i002p00247}.

\bibitem[{\textit{Lajeunesse et~al.}(2010)\textit{Lajeunesse, Malverti, and
  Charru}}]{Lajeunesse2010}
Lajeunesse, E., L.~Malverti, and F.~Charru (2010), Bed load transport in
  turbulent flow at the grain scale: Experiments and modeling, \textit{J.
  Geophys. Res.}, \textit{115}(F4), {F04,001}, \doi{10.1029/2009JF001628}.

\bibitem[{\textit{Martin et~al.}(2012)\textit{Martin, Jerolmack, and
  Schumer}}]{Martin2012}
Martin, R., D.~J. Jerolmack, and R.~Schumer (2012), The physical basis for
  anomalous diffusion in bed load transport, \textit{J. Geophys. Res.-earth},
  \textit{117}(F1), F01,018, \doi{10.1029/2011JF002075}.

\bibitem[{\textit{Nelson et~al.}(1995)\textit{Nelson, Shreve, McLean, and
  Drake}}]{nelson2005}
Nelson, J.~M., R.~L. Shreve, S.~R. McLean, and T.~G. Drake (1995), Role of
  near-bed turbulence structure in bed load transport and bed form mechanics,
  \textit{Water Resour. Res.}, \textit{31}(8), 2071--2086,
  \doi{10.1029/95WR00976}.

\bibitem[{\textit{Papanicolaou et~al.}(2002)\textit{Papanicolaou, Diplas,
  Evaggelopoulos, and Fotopoulos}}]{papa2002}
Papanicolaou, A.~N., P.~Diplas, N.~Evaggelopoulos, and S.~Fotopoulos (2002),
  Stochastic incipient motion criterion for spheres under various bed packing
  conditions, \textit{J. Hydraul. Eng.}, \textit{128}(4), 369--380,
  \doi{10.1061/(ASCE)0733-9429(2002)128:4(369)}.

\bibitem[{\textit{Radice et~al.}(2006)\textit{Radice, Malavasi, and
  Ballio}}]{radice2006}
Radice, A., S.~Malavasi, and F.~Ballio (2006), Solid transport measurements
  through image processing, \textit{Exp Fluids}, \textit{41}(5), 721--734,
  \doi{10.1007/s00348-006-0195-9}.

\bibitem[{\textit{Radice et~al.}(2009)\textit{Radice, Ballio, and
  Nikora}}]{Radice2009}
Radice, A., F.~Ballio, and V.~Nikora (2009), On statistical properties of bed
  load sediment concentration, \textit{Water Resour. Res.}, \textit{45}(6),
  n/a--n/a, \doi{10.1029/2008WR007192}.

\bibitem[{\textit{Ripley}(1976)}]{RipleyK}
Ripley, B. (1976), The second order analysis of stationary point processes,
  \textit{J. Appl. Probab.}, \textit{13}, 255--266.

\bibitem[{\textit{Roseberry et~al.}(2012)\textit{Roseberry, Schmeeckle, and
  Furbish}}]{Furbish2012(2)}
Roseberry, J., M.~Schmeeckle, and D.~Furbish (2012), A probabilistic
  description of the bed load sediment flux: 2. particle activity and motions,
  \textit{J. Geophys. Res.-earth}, \textit{117}(F3), F03,032,
  \doi{10.1029/2012JF002353}.

\bibitem[{\textit{Schmeeckle et~al.}(2001)\textit{Schmeeckle, Nelson, Pitlick,
  and Bennett}}]{Schmeeckle2001}
Schmeeckle, M., J.~Nelson, J.~Pitlick, and J.~Bennett (2001), Interparticle
  collision of natural sediment grains in water, \textit{Water Resour. Res.},
  \textit{37}(9), 2377--2391, \doi{10.1029/2001WR000531}.

\bibitem[{\textit{Schmeeckle et~al.}(2007)\textit{Schmeeckle, Nelson, and
  Shreve}}]{Schmeeckle2007}
Schmeeckle, M., J.~M. Nelson, and R.~Shreve (2007), Forces on stationary
  particles in near-bed turbulent flows, \textit{J. Geophys. Res.-earth},
  \textit{112}(F2), F02,003, \doi{10.1029/2006JF000536}.

\bibitem[{\textit{Seizilles et~al.}(2014)\textit{Seizilles, Lajeunesse,
  Devauchelle, and Bak}}]{Seizilles2014}
Seizilles, G., E.~Lajeunesse, O.~Devauchelle, and M.~Bak (2014), Cross-stream
  diffusion in bedload transport, \textit{Phys. Fluids}, \textit{26}(1),
  013302, \doi{http://dx.doi.org/10.1063/1.4861001}.

\bibitem[{\textit{Singh et~al.}(2009)\textit{Singh, Fienberg, Jerolmack, Marr,
  and Foufoula-Georgiou}}]{singh2009}
Singh, A., K.~Fienberg, D.~J. Jerolmack, J.~Marr, and E.~Foufoula-Georgiou
  (2009), Experimental evidence for statistical scaling and intermittency in
  sediment transport rates, \textit{J. Geophys. Res.}, \textit{114}(F1),
  F01,025, \doi{10.1029/2007JF000963}.

\bibitem[{\textit{Sun and Donahue}(2000)}]{sun2000}
Sun, Z., and J.~Donahue (2000), Statistically derived bedload formula for any
  fraction of nonuniform sediment, \textit{J. Hydraul. Eng.}, \textit{126}(2),
  105--111, \doi{10.1061/(ASCE)0733-9429(2000)126:2(105)}.

\bibitem[{\textit{Taylor}(1922)}]{taylor1922}
Taylor, G.~I. (1922), Diffusion by continuous movements, \textit{Proc. London
  Math. Soc.}, \textit{20}, 196–212.

\bibitem[{\textit{Turowski}(2010)}]{turowski2010}
Turowski, J.~M. (2010), Probability distributions of bed load transport rates:
  A new derivation and comparison with field data, \textit{Water Resour. Res.},
  \textit{46}(8), W08,501.

\bibitem[{\textit{Uhlenbeck and Ornstein}(1930)}]{Uhlenbeck1930}
Uhlenbeck, G.~E., and L.~S. Ornstein (1930), On the theory of the {B}rownian
  motion, \textit{Phys. Rev.}, \textit{36}, 823--841,
  \doi{10.1103/PhysRev.36.823}.

\bibitem[{\textit{Valyrakis et~al.}(2010)\textit{Valyrakis, Diplas, Dancey,
  Greer, and Celik}}]{Valyrakis2010}
Valyrakis, M., P.~Diplas, C.~Dancey, K.~Greer, and A.~Celik (2010), Role of
  instantaneous force magnitude and duration on particle entrainment,
  \textit{J. Geophys. Res.}, \textit{115}(F2), F02,006,
  \doi{10.1029/2008JF001247}.

\bibitem[{\textit{Wu and Chou}(2003)}]{wu2003}
Wu, F., and Y.~Chou (2003), Rolling and lifting probabilities for sediment
  entrainment, \textit{J. Hydraul. Eng.}, \textit{129}(2), 110--119,
  \doi{10.1061/(ASCE)0733-9429(2003)129:2(110)}.

\bibitem[{\textit{Wu and Yang}(2004)}]{wu2004}
Wu, F., and K.~Yang (2004), A stochastic partial transport model for mixed-size
  sediment: Application to assessment of fractional mobility, \textit{Water
  Resour. Res.}, \textit{40}(4), W04,501, \doi{10.1029/2003WR002256}.

\bibitem[{\textit{Yilmaz et~al.}(2006)\textit{Yilmaz, Javed, and
  Shah}}]{Yilmaz2006}
Yilmaz, A., O.~Javed, and M.~Shah (2006), Object tracking: A survey,
  \textit{ACM Comput. Surv.}, \textit{38}(4), \doi{10.1145/1177352.1177355}.

\end{thebibliography}

\end{document}